\documentclass[aps,twocolumn,amsmath,amssymb,showpacs,superscriptaddress,notitlepage,longbibliography]{revtex4-1}
\usepackage[colorlinks=true,linkcolor=blue,anchorcolor=red,citecolor=blue, urlcolor=blue]{hyperref}
\usepackage{bm}
\usepackage{graphicx}
\usepackage{color}
\usepackage{makecell}
\usepackage{multirow}

\begin{document}
\title{Probing electron-hole weights of an Andreev bound state by transient currents}
\author{Zhan Cao}
\email{caozhan@baqis.ac.cn}
\affiliation{Beijing Academy of Quantum Information Sciences, Beijing 100193, China}

\author{Gu Zhang}
\email{zhanggu@baqis.ac.cn}
\affiliation{Beijing Academy of Quantum Information Sciences, Beijing 100193, China}

\author{Hao Zhang}
\affiliation{State Key Laboratory of Low Dimensional Quantum Physics, Department of Physics, Tsinghua University, Beijing, 100084, China}
\affiliation{Beijing Academy of Quantum Information Sciences, Beijing 100193, China}
\affiliation{Frontier Science Center for Quantum Information, Beijing 100084, China}

\author{Wan-Xiu He}
\affiliation{Beijing Computational Science Research Center, Beijing 100193, China}

\author{Chuanchang Zeng}
\affiliation{Centre for Quantum Physics, Key Laboratory of Advanced Optoelectronic Quantum Architecture and Measurement (MOE), School of Physics, Beijing Institute of Technology, Beijing, 100081, China}

\author{Ke He}
\affiliation{State Key Laboratory of Low Dimensional Quantum Physics, Department of Physics, Tsinghua University, Beijing, 100084, China}
\affiliation{Beijing Academy of Quantum Information Sciences, Beijing 100193, China}
\affiliation{Frontier Science Center for Quantum Information, Beijing 100084, China}

\author{Dong E. Liu}
\email{dongeliu@mail.tsinghua.edu.cn}
\affiliation{State Key Laboratory of Low Dimensional Quantum Physics, Department of Physics, Tsinghua University, Beijing, 100084, China}
\affiliation{Beijing Academy of Quantum Information Sciences, Beijing 100193, China}
\affiliation{Frontier Science Center for Quantum Information, Beijing 100084, China}

\begin{abstract}
Andreev bound states (ABSs) are localized quantum states that contain both electron and hole components. They ubiquitously reside in inhomogeneous superconducting systems. Following theoretical analysis, we propose to probe the electron-hole weights of an ABS via a local tunneling measurement that detects the transient current under a steplike pulse bias. With our protocol, the ABS energy level can also be obtained from peaks of the Fourier spectrum of the transient current. Our protocol can be applied to detect robust zero-energy Majorana bound states (MBSs), which have equal electron-hole weights, in candidate platforms where local tunneling spectroscopy measurement is possible. In the 1D Majorana nanowire model, we numerically calculate the electron-hole weights for different types of low-energy bound states, including ABSs, quasi-MBSs, and MBSs.
\end{abstract}

%\date{\today}

\maketitle

\section{Introduction}
Andreev reflection~\cite{andreev1964thermal} is a unique transport mechanism that occurs at normal metal--superconductor interfaces. It converts incoming electrons (holes) into reflected holes (electrons), and generates Andreev bound states (ABSs)~\cite{sauls2018andreev} in inhomogeneous superconducting systems. ABSs widely exist in normal metal--superconductor tunnel junctions separated by single~\cite{deacon2010tunneling,dirks2011transport,lee2014spin,cao2017inelastic} or double~\cite{hofstetter2009cooper,herrmann2010carbon,hofstetter2011finite,schindele2012near,cao2015thermoelectric} quantum dots, Josephson junctions~\cite{beenakker1991Josephson,pillet2010andreev,de2010hybrid,martin2011josephson}, vortex cores in type-II superconductors~\cite{caroli1964bound,hess1989scanning,shore1989density,hess1990vortex}, surfaces of unconventional superconductors~\cite{tanaka1995theory,laube2000spin,deutscher2005andreev}, etc. As ABSs arise from Andreev reflections, they generically contain both electron and hole components~\cite{de1963elementary,kulik1969macroscopic}, indicated by the corresponding weights $w_e$ and $w_h$, respectively. Especially, these two weights equal ($w_e = w_h$) for an exotic zero-energy ABS, i.e., the Majorana bound state (MBS)~\cite{read2000paired,kitaev2001unpaired,prada2020andreev} at the edges of topological superconductors~\cite{alicea2012new}. The electron-hole superposition of both ABSs and MBSs uniquely distinguishes them from most other quasi-particle states.

It is of general interest to detect and control ABSs in different systems with tuning knobs such as gate voltages and external magnetic fields. In experiments, one can routinely obtain discrete energy levels of ABSs via the local tunneling spectroscopy measurement realized by a metallic probe with a dc bias voltage. This simple spectroscopy measurement, however, detects the product $w_e w_h$ of an ABS~\cite{schindele2014nonlocal}, and is thus unable to reveal the individual values of $w_e$ and $w_h$. The existing protocols of measuring $w_e$ and $w_h$ of an ABS rely on the Coulomb-blockade conductance peaks in superconducting islands~\cite{hansen2018probing} or the nonlocal conductance in three-terminal setups~\cite{danon2020nonlocal}. Recently, both protocols have been applied to detect MBSs (owning $w_e = w_h$ ) that are believed to exist in semiconductor-superconductor hybrid nanowire devices~\cite{o2018hybridization,shen2018parity,menard2020conductance,poschl2022nonlocal}. Though these two protocols are well designed for nanowire devices, their implementations could be potentially rather complicated and difficult in other possible MBS-hosting systems, e.g., topological insulator-superconductor heterostructures~\cite{xu2015experimental,sun2016majorana} and iron-based superconductors~\cite{zhang2018observation,wang2018evidence}. It is thus rewarding to design a feasible protocol that can probe $w_e$ and $w_h$ of an ABS in general systems. As aforementioned, local tunneling spectroscopy is a widely used technique, and it can be easily applied to most superconducting systems. Is it possible to design an extended version of the local tunneling spectroscopy that can detect the values of $w_e$ and $w_h$?

In this work, we propose that $w_e$ and $w_h$ of an ABS are experimentally accessible via a local tunneling current measurement using a steplike pulse bias. More specifically, $w_e$ and $w_h$ of an ABS can be extracted from the time evolution of the transient current; the ABS energy level can be obtained from the characteristic peak positions of the Fourier spectrum of the transient current. The knowledge of the electron-hole weights and energy level of an ABS can be applied to distinguishing among zero-energy ABSs, quasi-MBSs \cite{vuik2019reproducing,moore2018two,moore2018quantized,stanescu2019robust,cao2019decays,kells2012near,liu2017Andreev,fleckenstein2018decaying}, and MBSs, which is yet highly demanded in the current research of MBSs~\cite{prada2020andreev,cao2022recent}. Based on the 1D Majorana nanowire model~\cite{oreg2010helical,lutchyn2010majorana}, we numerically calculate the spatial profiles of the wave functions and electron-hole weights $w_e$, $w_h$ for ABSs, quasi-MBSs, and MBSs. We want to point out that the obtained distinct features suffice to distinguish MBSs from ABSs and quasi-MBSs. Importantly, we anticipate those features as experimentally accessible, following recent advances in ultrafast electronic transport techniques~\cite{mciver2020light,torre2021colloquium,gutzler2021light}.

The remainder of this paper is organized as follows. In Sec.~\ref{secII}, we introduce the model Hamiltonian, explain our protocol by a simple semi-classical method, and establish an exact time-dependent current formula. In Sec.~\ref{results}, we first show the numerical results of time-dependent current mediated by ABSs with different electron-hole weights. We further present the calculated energy spectra, wave functions, and electron-hole weights of different types of bound states in the 1D Majorana nanowire model. Finally, we give a brief summary in Sec.~\ref{summary}. Calculation details are included in the Appendices.

\begin{figure}[t!]
\centering
\includegraphics[width=0.95\columnwidth]{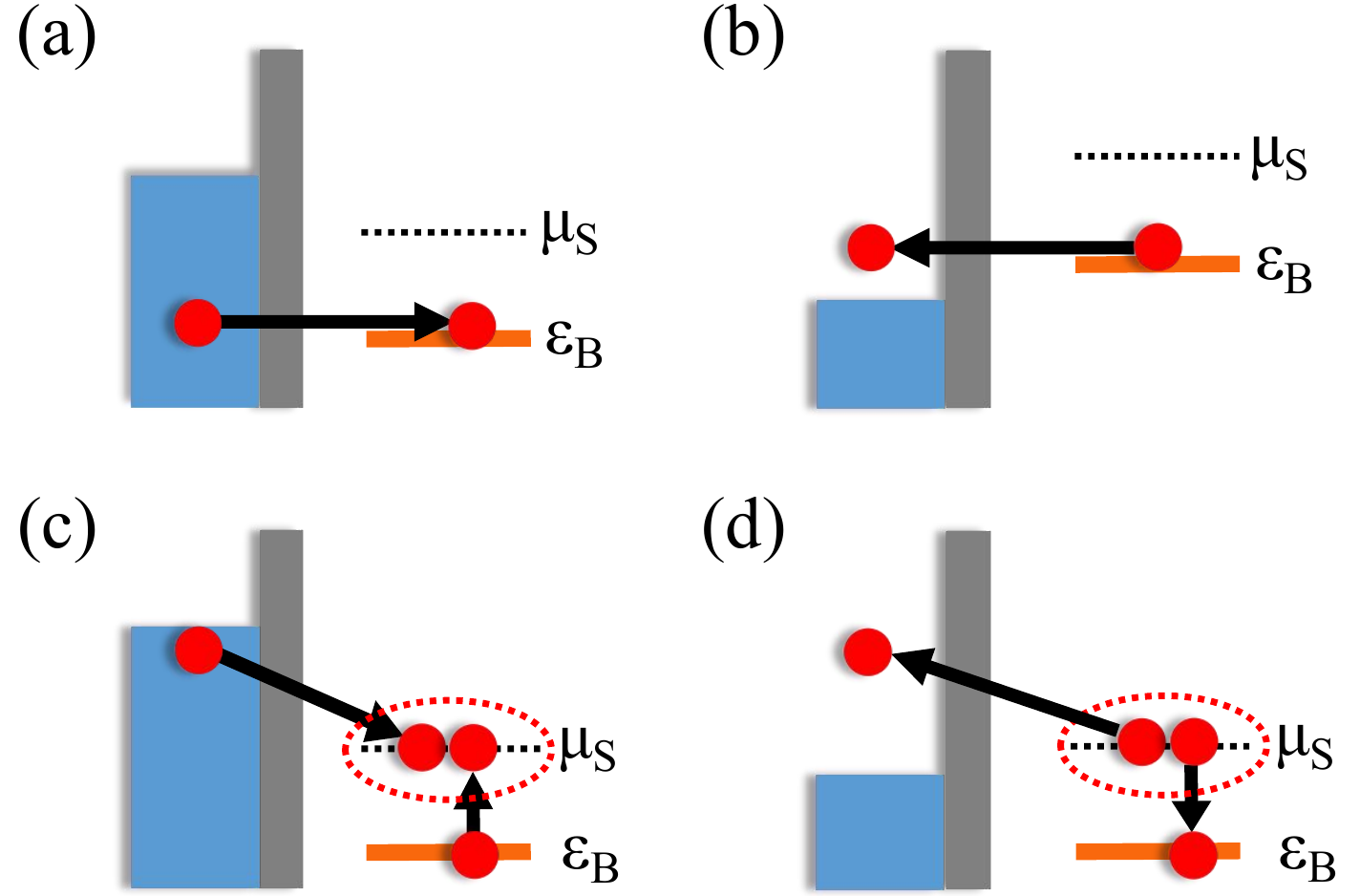}
\caption{Schematics of normal [(a) and (b)] and Andreev [(c) and (d)] sequential tunnelings between a metallic probe (blue rectangle) and an ABS with energy $\varepsilon_B$ across a tunneling barrier (gray rectangle). The two electrons enclosed in an ellipse represent a Cooper pair condensed at the Fermi level $\mu_S$ of the superconductor. See text for the description of these tunneling processes.}\label{Fig:illu}
\end{figure}

\section{Model and formalism}\label{secII}
\subsection{Effective Hamiltonian}\label{secIIA}
The system we consider can be modeled by the effective Hamiltonian
\begin{eqnarray}
H&=&H_P+H_B+H_T,\label{Htotal}\\
H_P&=&\sum_{k\sigma}[\varepsilon_{k}+U(t)]c_{k\sigma}^{\dag}c_{k\sigma},\\
H_{B}&=&\varepsilon_{B}c_{B}^{\dag}c_{B},\label{H_ABS}\\
H_{T}&=&\lambda\sum_{k\sigma}c_{k\sigma}^{\dag}(  u_{\sigma}c_{B}+v_{\sigma}c_{B}^{\dag})+\textrm{H.c.}.\label{Htunneling}
\end{eqnarray}
Here, $H_P$ describes a metallic probe with energy dispersion $\varepsilon_k$ under a time-dependent bias voltage $U(t)$~\cite{jauho1994time,maciejko2006time}. The probe, which can be an electrode or a STM tip, is assumed to be a metal with a constant density of states $\rho$. $H_{B}$ describes an ABS with energy $\varepsilon_B$ smaller than the superconducting gap. We choose the Fermi level $\mu_S$ of the superconductor as the reference energy. The wave function of the ABS is characterized by the position ($x$) and spin ($\sigma$) dependent Bogoliubov-de Gennes (BdG) amplitudes $u_\sigma(x)$ and $v_\sigma(x)$. The position-dependent electron and hole weights of the ABS can be defined as $w_e(x)=\sum_\sigma|u_\sigma(x)|^2$ and $w_h(x)=\sum_\sigma|v_\sigma(x)|^2$, respectively. The tunneling Hamiltonian $H_{T}$ describes the local coupling between the probe and the ABS with a constant tunneling amplitude $\lambda$ \cite{hansen2018probing,liu2022universal}. The coupling induced level broadening is $\Gamma=2\pi\rho|\lambda|^2$. In $H_T$, we use $u_\sigma$ and $v_\sigma$ (with $x$ being omitted) to denote the local BdG amplitudes at the position nearest to the probe. For simplicity, one can redefine the parameter $\lambda$ in $H_T$ to impose a normalization condition $\sum_\sigma(|u_\sigma|^2+|v_\sigma|^2)=1$, i.e., $w_e+w_h=1$. We also assume $\varepsilon_B\le 0$ throughout the work, as the results of $\varepsilon_B > 0$ cases can be immediately obtained via a particle-hole transformation.

\subsection{Semi-classical rate equation analysis}\label{secIIB}
Our protocol of probing $w_e$ and $w_h$ of an ABS via a local tunneling current measurement is established by the semi-classical rate equation (RE) approach~\cite{bonet2002solving,koch2004thermopower,koch2006theory,he2020performance,bruus2004many}. Briefly, the time-dependent occupation [$P_1(t)$] and inoccupation [$P_0(t)$] probabilities of an ABS obey the normalization requirement $P_0(t)+P_1(t)=1$ and the RE
\begin{equation}
\frac{dP_1(t)}{dt}=-\gamma_{1\rightarrow 0}(t)P_1(t)+\gamma_{0\rightarrow 1}(t)P_0(t),
\end{equation}
where $\gamma_{1\rightarrow0}$ ($\gamma_{0\rightarrow 1}$) refers to the transition rate from the occupied (unoccupied) to the unoccupied (occupied) state. As an ABS consists of electron and hole components, its communication with the probe contains both the normal ($N$) and Andreev ($A$) sequential tunnelings, as illustrated in Fig.~\ref{Fig:illu}. Specifically, when the ABS is unoccupied and $\varepsilon_B<U(t)$, an electron with energy $\varepsilon_B$ in the probe can tunnel into the ABS with a rate $\gamma_{0\rightarrow1}^N(t)$ [Fig.~\ref{Fig:illu}(a)]. Inversely, an electron can tunnel from the ABS into the probe with a transition rate $\gamma_{1\rightarrow0}^N(t)$ if the ABS is occupied and $\varepsilon_B>U(t)$ [Fig.~\ref{Fig:illu}(b)]. When the ABS is occupied and $U(t)>-\varepsilon_B$, an Andreev tunneling with a transition rate $\gamma_{1\rightarrow0}^A(t)$ could convert an electron with energy $-\varepsilon_B$ in the probe and the electron on the ABS into a Cooper pair [Fig.~\ref{Fig:illu}(c)]. Similarly, the opposite process, i.e., Cooper pair splitting, with a transition rate $\gamma_{0\rightarrow1}^A(t)$ is energetically allowed if the ABS is unoccupied and $U(t)<-\varepsilon_B$ [Fig.~\ref{Fig:illu}(d)]. As a result, the rates in the RE are $\gamma_{0\rightarrow1}(t)=\gamma^N_{0\rightarrow1}(t)+\gamma^A_{0\rightarrow1}(t)$ and $\gamma_{1\rightarrow0}(t)=\gamma^N_{1\rightarrow0}(t)+\gamma^A_{1\rightarrow0}(t)$.
Following Fermi's golden rule \cite{bruus2004many},
\begin{eqnarray}
\gamma_{0\rightarrow1}^{N}(t)&=&\frac{\Gamma}{\hbar}f(  \varepsilon_{B}-U(t))  w_{e},\\ \gamma_{1\rightarrow0}^{N}(t)&=&\frac{\Gamma}{\hbar}[  1-f(\varepsilon_{B}-U(t))  ]  w_{e},\\
\gamma_{0\rightarrow1}^{A}(t)&=&\frac{\Gamma}{\hbar}[  1-f(  -\varepsilon_{B}-U(t))  ]w_{h},\\
\gamma_{1\rightarrow0}^{A}(t)&=&\frac{\Gamma}{\hbar}f(  -\varepsilon_{B}-U(t))  w_{h},
\end{eqnarray}
where $f(\varepsilon)=1/(e^{\varepsilon/k_B T}+1)$ is the Fermi-Dirac distribution function. The RE method can capture the major features of our system at high enough temperatures ($k_BT \gtrsim \Gamma$).

For the simplest situation, we consider a steplike pulse bias $U(t)=\Theta(t)V$, which has been widely studied for transient transports in normal~\cite{jauho1994time,stefanucci2004time,maciejko2006time,souza2007transient,zheng2008adynamic,perfetto2010correlation,wang2010transient,tuovinen2013time,tuovinen2014time,gaury2015ac,antipov2016voltage,ridley2021quantum} and superconducting ~\cite{xing2007response,tuovinen2016time,tuovinen2019distinguishing} systems. Before applying the bias, the ABS has the equilibrium occupation $P_{1}(t\le 0)=f( \varepsilon_{B})$, which is obtained by setting $dP_1(t)/dt=0$ in the RE. After the sudden switch-on of a constant bias, the solution of the RE instead becomes $P_{1}(  t>0)=f(\varepsilon_B-V)w_e+f(\varepsilon_B+V)w_h
+(\delta f_e w_e-\delta f_h w_h)e^{-\Gamma t/\hbar}$, where $\delta f_{e(h)}=f(\pm \varepsilon_B)-f(\pm \varepsilon_B-V)$ refers to the non-equilibrium bias induced modification of the distribution. The local tunneling current, $I(  t)=-e[  \gamma_{1\rightarrow0}^{N}(  t)  -\gamma_{1\rightarrow0}^{A}(  t)  ]  P_{1}(  t)
+e[  \gamma_{0\rightarrow1}^{N}(t)  -\gamma_{0\rightarrow1}^{A}(  t)  ]  [1-P_{1}(t)]$, then equals
\begin{eqnarray}
I(t)&=&\Theta(t)[I_\textrm{st}+I_\textrm{trans}(t)],\label{It_re}\\
I_\textrm{st}  &=&2e\frac{\Gamma}{\hbar}[f(\varepsilon_B-V)-f(\varepsilon_B+V)]w_e w_h,\label{Ist_re}\\
\hspace{-0.3cm}I_\textrm{trans}(t)  &=&-e\frac{\Gamma}{\hbar}(w_e-w_h)(\delta f_e w_e-\delta f_h w_h)e^{-\Gamma t/\hbar},\label{Itrans_re}
\end{eqnarray}
where $I_\textrm{st}$ and $I_\textrm{trans}(t)$ are steady-state and transient currents, respectively.

Equations~\eqref{It_re}--\eqref{Itrans_re} lead to one of our major conclusions: the individual values of $w_e$ and $w_h$, which are inaccessible from the steady-state current $I_\textrm{st}$, can be obtained by measuring the transient current $I_\textrm{trans}(t)$ at temperatures $k_BT \gtrsim \Gamma$. Specifically, the values of $w_e$ (or $w_h=1-w_e$) and $\Gamma$ can be readily obtained by fitting the measured time evolution of the transient current with our theoretical expressions Eqs.~\eqref{It_re}--\eqref{Itrans_re}, after knowing $V$, $T$ and $\varepsilon_B$.
The former two ($V$ and $T$) are experimental knob parameters that can be directly read out. The value of $\varepsilon_B$ can be obtained from the Fourier spectrum of the transient current at temperatures $k_BT < \Gamma$, as will be shown in Sec.~\ref{results}.

\subsection{Nonequilibrium Green's function formalism}\label{secIIC}
To support the RE analysis, we calculate the current with the exact nonequilibrium Green's function (NEGF) technique~\cite{jauho1994time,haug2008quantum}. With our chosen steplike pulse bias, the exact steady-state current is (see Appendix \ref{appa})
\begin{equation}
I_{\textrm{st}}=\frac{e}{2h}\int d\varepsilon[  f(\varepsilon-V)  -f(  \varepsilon+V) ]T(\varepsilon),\label{Ist}
\end{equation}
with the transmission probability
\begin{equation}
T(\varepsilon)=\sum_{\eta=e,h}\textrm{Tr}\big[\mathbf{\Gamma}_{\eta}\mathbf{G}^{r}(  \varepsilon)\mathbf{\Gamma}_{\bar\eta}\mathbf{G}^{a}(  \varepsilon)\big],\label{TE}
\end{equation}
where $\mathbf{G}^{r(a)}(\varepsilon)=1/\big(\varepsilon-\mathbf{H}_S^{r(a)}\big)$ is the dressed Green's function of the ABS with $\mathbf{H}_S^{r(a)}=\varepsilon_B \sigma_z \mp i (\mathbf{\Gamma}_{e}+\mathbf{\Gamma}_{h})/2$ ($\sigma_z$ is the Pauli matrix in Nambu space) and
\begin{equation}
\mathbf{\Gamma}_{e}=\Gamma\left(
\begin{array}
[c]{cc}
w_{e} &w_{eh}\\
w^\ast_{eh} &w_{h}
\end{array}
\right),~\mathbf{\Gamma}_{h}=\Gamma\left(
\begin{array}
[c]{cc}
w_{h} &w_{eh}\\
w^\ast_{eh} &w_{e}
\end{array}
\right),
\end{equation}
with weight $w_{eh}=\sum_\sigma u^\ast_\sigma v_\sigma$, which is irrelevant in the RE analysis. Clearly, $I_\textrm{st}$ and the differential conductance, $G\equiv dI_\textrm{st}/dV$, obey the symmetries $I_{\textrm{st}}(V)=-I_{\textrm{st}}(-V)$ and $G(V)=G(-V)$, which has been explained by the particle-hole symmetry and the unitarity of the scattering matrix of a normal metal--superconductor junction \cite{melo2021conductance}.
The explicit expression of the transmission probability in Eq.~\eqref{TE} is
\begin{eqnarray}
\hspace{-0.5cm}T(\varepsilon)  =\frac{16\Gamma^{2}[  4 \varepsilon^{2}p_{+}+p_{-}(  \Gamma^{2}q_{-}+4\varepsilon_{B}^{2})  ]  }{\Gamma^{4}q_{-}^{2}+16(  \varepsilon_{B}^{2}-\varepsilon^{2}) ^{2}+8\Gamma^{2}(  \varepsilon^{2}q_{+}+\varepsilon_{B}^{2}q_{-})  },
\label{eq:steady_state_transmission}
\end{eqnarray}
where $p_{\pm}=w_{e}w_{h}\pm |w_{eh}|^{2}$ and $q_{\pm}=1\pm4 |w_{eh}|^{2}$. When $V\gg k_B T,~\Gamma,~|\varepsilon_B|$, the steady-state current in Eq.~\eqref{Ist} becomes $I_\textrm{st}=2e\Gamma w_ew_h/\hbar$, consistent with Eq.~\eqref{Ist_re} of the RE analysis. As only the product of $w_e$ and $w_h$ appear in Eq.~\eqref{eq:steady_state_transmission}, their individual values are inaccessible from $I_\textrm{st}(V)$ and $G(V)$. At zero temperature, for a zero-energy MBS owning $u_\uparrow/v_\uparrow=u_\downarrow/v_\downarrow=e^{i\phi}$ ($\phi$ is an arbitrary real number), i.e., $\varepsilon_B=0$ and $w_e=w_h=|w_{eh}|=0.5$, Eqs.~\eqref{Ist} and \eqref{eq:steady_state_transmission} lead to the well-known quantized zero-bias conductance, i.e., $G(0)=2e^2/h$~\cite{law2009majorana,flensberg2010tunneling}.

The exact transient current $I_\textrm{trans}(t)$ is a sum of the following two parts (see Appendix \ref{appa})
\begin{eqnarray}
I_{1}(t)&=&\frac{e}{2h}\sum_{\eta=e,h}\int d\varepsilon f(s_\eta\varepsilon)\textrm{Tr}\big[is_\eta V\mathbf{\Gamma}_{\eta}e^{i(  \varepsilon+ s_\eta V-\mathbf{H}_{S}^{r})  t}\notag\\
&&\times \mathbf{G}^{r}(  \varepsilon)  \mathbf{G}^{r}(\varepsilon+s_\eta V) +\textrm{H.c.}\big],\label{I1}\\
I_{2}(  t)&=&\frac{e}{2h}\sum_{\eta=e,h}\int d\varepsilon f(\varepsilon) \textrm{Tr}\Big\{\big[-s_\eta V (\mathbf{\Gamma}_{e}-\mathbf{\Gamma}_{h})\notag\\
&&\times e^{i(  \varepsilon+s_\eta V-\mathbf{H}_{S}^{r})  t}\mathbf{G}^{r}(  \varepsilon)  \mathbf{B}_\eta(  \varepsilon)  +\textrm{H.c.}\big]\notag\\
&&\hspace{-0.7cm}-V^{2}(\mathbf{\Gamma}_{e}-\mathbf{\Gamma}_{h})e^{-i\mathbf{H}_{S}^{r}t}\mathbf{G}^{r}(\varepsilon)  \mathbf{B}_\eta(  \varepsilon)\mathbf{G}^{a}(  \varepsilon)    e^{i\mathbf{H}_{S}^{a}t}\Big\},\label{I2}
\end{eqnarray}
where  $s_{e(h)}=\pm 1$ and $\mathbf{B}_\eta(  \varepsilon)=\mathbf{G}^{r}(  \varepsilon+s_\eta V)  \mathbf{\Gamma}_{\eta}\mathbf{G}^{a}(  \varepsilon+s_\eta V)$. The matrix $\mathbf{H}_S^{r(a)}$ is non-Hermitian with eigenvalues
\begin{eqnarray}
E_{1,2}^{r}=\pm\sqrt{\varepsilon_B^2-w_{eh}^2\Gamma^2}- i\Gamma/2,\ E_{1,2}^a=E_{1,2}^{r*}.
\label{E12r}
\end{eqnarray}
The traces in Eqs.~\eqref{TE}, \eqref{I1}, and \eqref{I2} are evaluated based on the eigenbasis of $\mathbf{H}_S^{r(a)}$ (see Appendix \ref{appa}). As $\textrm{Im}E_{1,2}^{r}\le 0$ and $\textrm{Im}E_{1,2}^{a}\ge 0$, the exponential factors $\exp(-i\mathbf{H}_{S}^{r}t)$ and $\exp(i\mathbf{H}_{S}^{a}t)$ of $I_1(t)$ and $I_2(t)$ die out in the long-time limit $t\gg\hbar/\Gamma$, in agreement with their transient feature obtained from the RE analysis. Additionally, the transient current $I_1 + I_2$ breaks the symmetry respected by $I_\textrm{st}$, i.e., $I(t,V)\neq -I(t,-V)$. In effect, it is this particle-hole asymmetry in the nonequilibrium transient dynamics that enables the measurement of $w_e$ and $w_h$ of an ABS using the transient transport.

\begin{figure}[t!]
\centering
\includegraphics[width=\columnwidth]{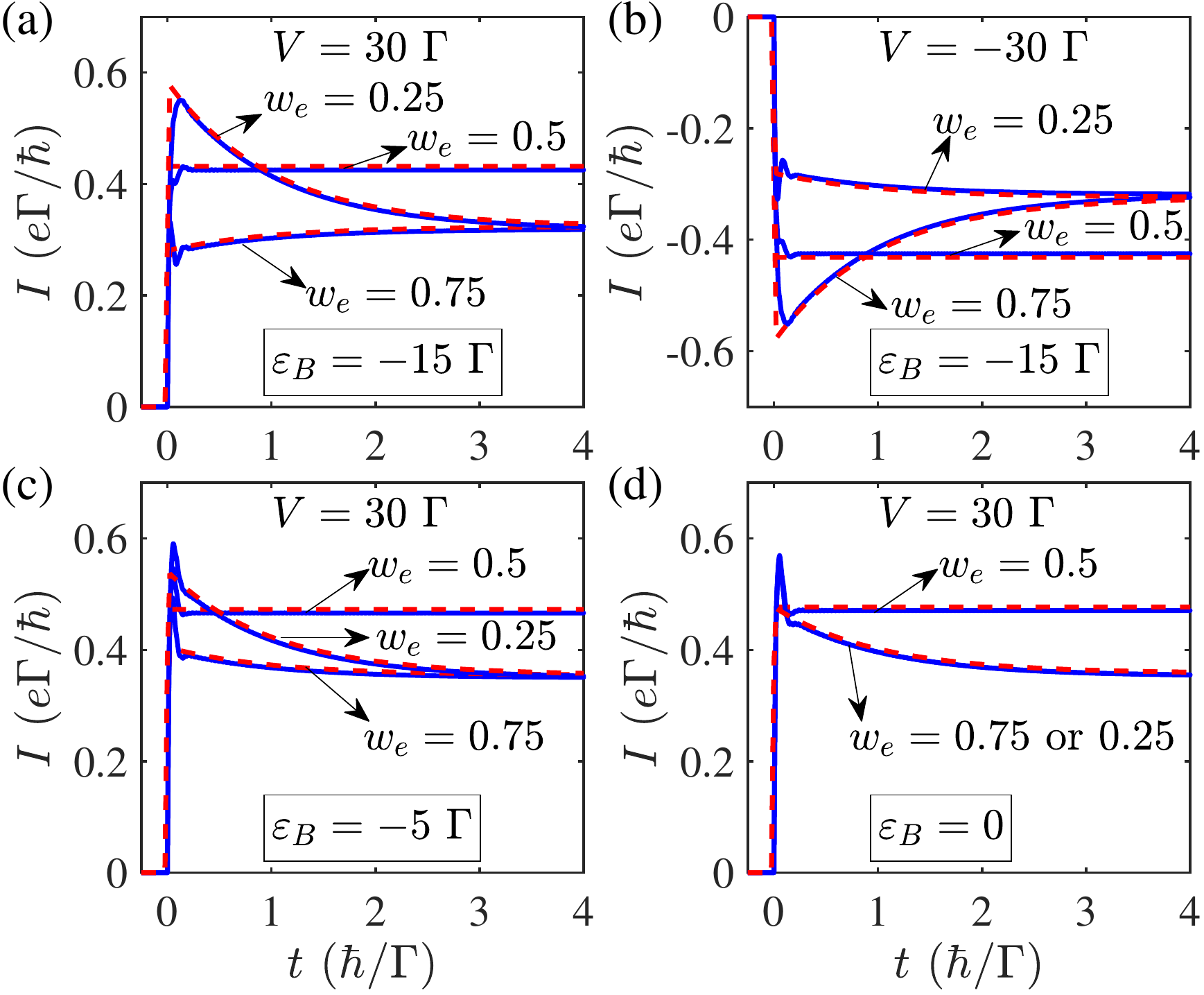}
\caption{Time evolution of the tunneling current at $k_BT=8~\Gamma$ under a steplike pulse bias $U(t) = \Theta(t)V$. The blue solid and red dashed curves are obtained respectively using the NEGF and the RE methods. The bias voltage $V=30~\Gamma$ in (a), (c), and (d), while $V=-30~\Gamma$ in (b). Different ABS energy levels $\varepsilon_B$ and electron weights $w_e$ have been considered. We choose $w_{eh}=0$ in the NEGF calculation.}\label{Fig:comparison}
\end{figure}

\section{Numerical results}\label{results}
\subsection{Time-dependent current mediated by an ABS}\label{resultA}
Figure \ref{Fig:comparison} compares the time evolution of current $I(t)$ calculated with the RE (red dashed lines) and the NEGF (blue solid lines), respectively, at high temperatures ($k_B T\gtrsim\Gamma$). Different parameter sets are considered for three regimes, i.e., $\Gamma< k_B T<|\varepsilon_B|$ [Figs.~\ref{Fig:comparison}(a) and \ref{Fig:comparison}(b)], $\Gamma<|\varepsilon_B|<k_B T$ [Fig.~\ref{Fig:comparison}(c)], and $|\varepsilon_B|<\Gamma< k_B T$ [Fig.~\ref{Fig:comparison}(d)], as given in the corresponding panels in Fig.~\ref{Fig:comparison}. We keep $|V|$ as the largest energy scale to enable the ABS-mediated sequential tunnelings mentioned in Sec.~\ref{secIIA}. In Fig.\,\ref{Fig:comparison}, we observe remarkable agreement of the $I(t)$ from the two methods, except for the initial weak current oscillations, for all three regimes under consideration. Note that, the coherent current oscillations captured by the NEGF is completely ignored by the RE as it involves only incoherent sequential tunnelings. In Fig.~\ref{Fig:comparison}(a) ($V=30~\Gamma$ and $\varepsilon_B=-15~\Gamma$), the currents flowing into ABSs with different electron weights $w_e=0.75$ and $w_e=0.25$, which represent particle-hole conjugate situations, evolve oppositely (upward/downward) and exponentially (at a rate $\Gamma/\hbar$) into the identical steady-state value. When the bias reverses its sign [Fig.~\ref{Fig:comparison}(b)], the currents instead becomes $I(t,V,w_e)=-I(t,-V,1-w_e)$, which is evident from Eqs.~\eqref{It_re}--\eqref{Itrans_re}. This modified symmetry, which becomes now weight-dependent, is indeed the reason one can extract $w_e$ and $w_h$ from the transient current.

The current evolution also depends on the ABS energy level. For $\varepsilon_B=-5~\Gamma$ [Fig.~\ref{Fig:comparison}(c)] and $\varepsilon_B=0$ [Fig.~\ref{Fig:comparison}(d)], the current curves of $w_e=0.25$ and $w_e=0.75$ both show a downward trend, in contrast to those in Figs.~\ref{Fig:comparison}(a) and \ref{Fig:comparison}(b). Moreover, we highlight that the difference between current curves of $w_e=0.75$ and $w_e=0.25$ decreases with reducing $|\varepsilon_B|$. The difference finally disappears for $\varepsilon_B = 0$, as the system becomes perfectly particle-hole symmetric. As another important feature, in all four panels, current curves of $w_e = 0.5$ arrive at their steady-state values immediately after the switch-on of the pulse bias, due to the absence of transient current component as addressed by Eq.~\eqref{Itrans_re}. By contrast, the current curves of $w_e=0.75$ and $w_e=0.25$ have an apparent exponential evolution in time. We want to stress that, as MBSs have $w_e = 0.5$ following its definition, the distinguishing feature of the $w_e = 0.5$ current curves is potentially a strong signature to identify MBSs.

\begin{figure}[t!]
\centering
\includegraphics[width=0.95\columnwidth]{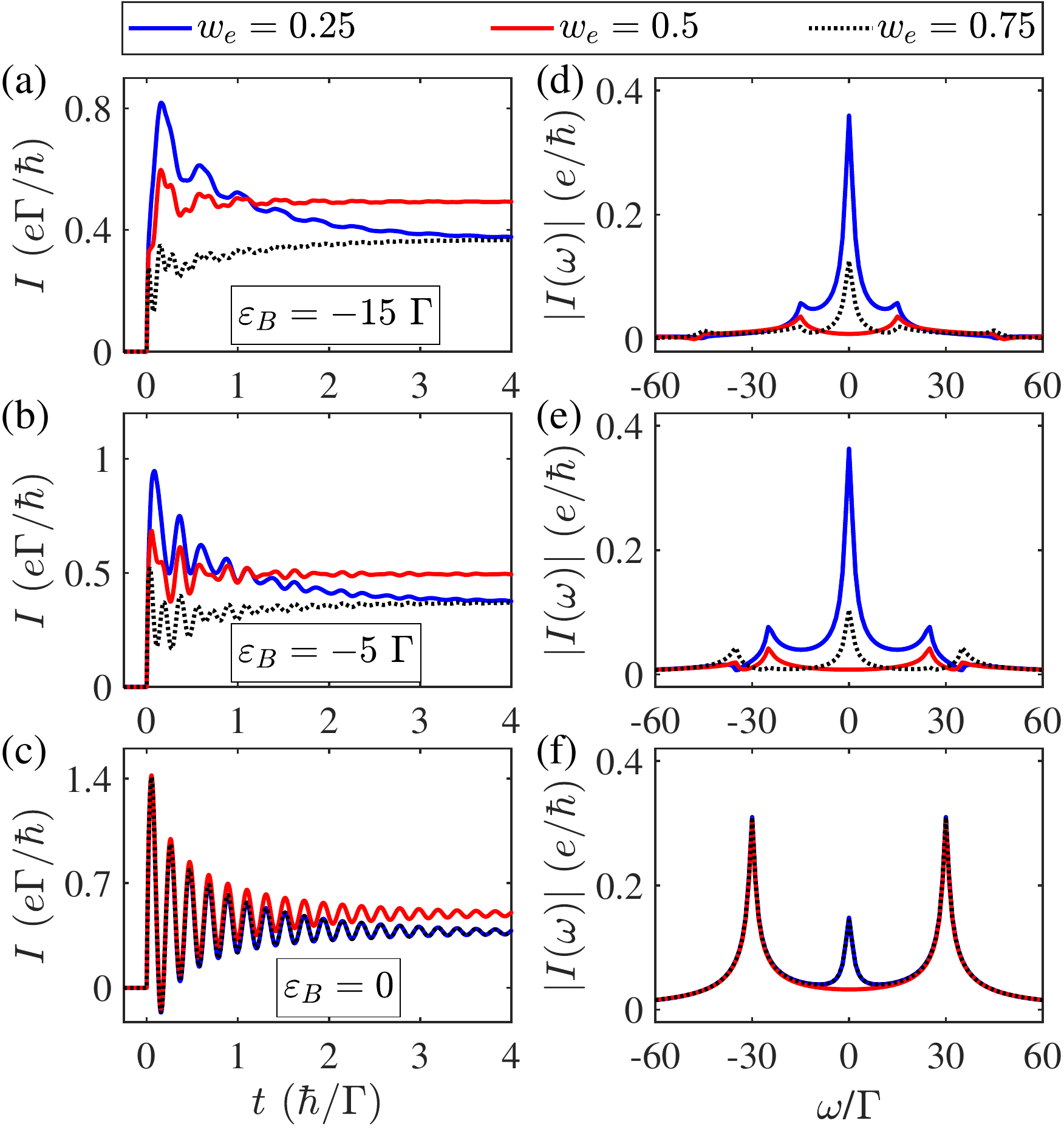}
\caption{[(a)--(c)] Time evolution of the tunneling current at $k_BT=0$ under a steplike pulse bias $U(t)=\Theta(t)V$ with $V=30~\Gamma$. Current curves are obtained with the NEGF method for ABSs with different energy levels $\varepsilon_B$ and electron weights $w_e$. [(d)--(f)] The corresponding modulus of the Fourier spectrum of the transient current in the left panel. The finite-frequency spectrum peaks locate at $|\omega_\pm|\approx |V|\pm|\varepsilon_B|$. The zero-frequency peak is absent if $w_e=0.5$, for all values of $\varepsilon_B$. Note that the lines of $w_e=0.75$ and $w_e=0.25$ coincide in (c) and (f), where $\varepsilon_B = 0$. We choose $w_{eh}=0$ as in Fig.~\ref{Fig:comparison}.}
\label{Fig:NEGF}
\end{figure}

The ABS energy level $\varepsilon_B$ can be obtained from the Fourier spectrum of the transient current $I(\omega)=\int_0^\infty dt I_\textrm{trans}(t)e^{i\omega t}$ at relatively low temperatures $k_B T < \Gamma$. In this limit, we choose $T=0$ in numerical calculations for simplicity. As shown in Figs.~\ref{Fig:NEGF}(a)--\ref{Fig:NEGF}(c), $I(t)$ displays oscillating and decaying features that originate from the imaginary and real parts, respectively, of the exponential factors in $I_1(t)$ and $I_2(t)$ [see Eqs.~\eqref{I1} and \eqref{I2}].
The current oscillations induced by quantum coherence are suppressed at high temperatures (see Fig.~\ref{Fig:comparison}). The modulus of the corresponding Fourier spectra $|I(\omega)|$ are shown in Figs.~\ref{Fig:NEGF}(d)--\ref{Fig:NEGF}(f).
The spectra have two types of characteristic peak: a Lorentzian peak at $\omega_0=0$ with a width $\Gamma$ and four logarithmic-type peaks at $|\omega_{\pm}|=|V|\pm \textrm{Re}E^r_{1}$ (see Appendix \ref{appc}). Physically, the zero-frequency peak originates from the exponential current evolution captured by both the RE and the $I_2(t)$ from the NEGF, while the finite-frequency peaks are associated with current oscillations induced by coherent electron tunnelings between the probe and the ABS. For a weak coupling strength $\Gamma$, $\textrm{Re}E_1^r$ approximately equals $-\varepsilon_B$ [see Eq.~\eqref{E12r}]. As a result, the ABS energy level $\varepsilon_B$ can then be obtained from the finite-frequency peaks of the spectrum, following $\varepsilon_B = |V|-|\omega_+|$.

Notably, in Figs.~\ref{Fig:NEGF}(d)--\ref{Fig:NEGF}(f), the zero-frequency peak is absent if $w_e=0.5$, for all chosen values of $\varepsilon_B$.
Indeed, we have $\mathbf{\Gamma}_e=\mathbf{\Gamma}_h$ when $w_e=0.5$, leading to a vanishing current $I_2(t)$ [see Eq.~\eqref{I2}]. Thus, the zero-frequency peak in the Fourier spectrum that depends on the exponential suppression of $I_2(t)$ disappears. Correspondingly, at low temperatures ($k_B T<\Gamma$), $I(t)$ oscillates around the steady-state value as shown by the red lines in Figs.~\ref{Fig:NEGF}(a)--\ref{Fig:NEGF}(c); at high temperatures ($k_B T\gtrsim \Gamma$), $I(t)$ jumps immediately to its steady-state value when the bias is suddenly switched on [see Figs.~\ref{Fig:comparison}(a)--\ref{Fig:comparison}(d) for $w_e=0.5$].

As discussed above, there are two relevant time scales of the transient current: the exponential decaying time scale $\tau_1 = \hbar/\Gamma$ and the period of the coherent oscillation $\tau_2 = \hbar/(|V|+|\varepsilon_B|)$.
Briefly, $\tau_1$ and $\tau_2$ refer to the time scales in which the weights and the energy level of the ABS become experimentally accessible, respectively. To estimate the typical values of $\tau_1$ and $\tau_2$, we take $\Delta=0.3$ meV for the commonly used conventional superconductor Al. Within our protocol, $|\varepsilon_B|<|V|<\Delta$ is required, allowing for only sequential tunnelings mediated by the ABS. We thus assume $\Gamma=0.01$ meV, $|\varepsilon_B|=0.05$ meV, and $|V|=0.1$ meV, and obtain $\tau_1=66$ ps and $\tau_2=4.4$ ps. The associated tunneling current is of the order of $1$ nA, which is large enough to be measured in experiments. Experimentally, the smallness of $\tau_2$ might pose a challenge for obtaining $|\varepsilon_B|$ from the coherent oscillation of the current evolution. However, the value of $|\varepsilon_B|$ can alternatively be obtained by the standard tunneling spectroscopy measurement. In realistic experiments, the value of $\Gamma$ can be further reduced by one order or more. In addition, the bias $V$ can also be smaller when detecting a near-zero-energy ABS with a small $\varepsilon_B$, as long as $|V| > |\varepsilon_B| $. In these cases, the values of $\tau_1$ and $\tau_2$ become larger, making the detection of ABS features more feasible. Therefore, we anticipate the experimental accessibility of the current evolution features displayed in Figs.~\ref{Fig:comparison} and \ref{Fig:NEGF} by the sub-picosecond measurement techniques, which have been achieved recently in ultrafast electronic transport measurements \cite{mciver2020light,torre2021colloquium,gutzler2021light}.

\subsection{Numerical simulations of the 1D Majorana nanowire model}\label{resultB}
For a generic ABS, its electron weight $w_e$ and energy level $\varepsilon_B$ are normally unrelated. By contrast, an exotic MBS requires both zero energy ($\varepsilon_B=0$) and specific weights ($w_e=w_h=\vert w_{eh}\vert=0.5$ as mentioned in Sec.~\ref{secIIC}). In our protocol, the weight $|w_{eh}|$ is inaccessible as it is absent in Eqs.~\eqref{It_re}--\eqref{Itrans_re}.
Strictly, the value of $|w_{eh}|$ can be obtained from the heights and line shapes of conductance peaks [see Eqs.~\eqref{Ist} and \eqref{eq:steady_state_transmission}]. It is however much more complicated than the readouts of $w_e$ and $w_h$ from the transient current [see Eq.~(\ref{Itrans_re})]. Nevertheless, we infer that $|w_{eh}|=0.5$ is implicit when both $w_e=w_h=0.5$ and $\varepsilon_B=0$ persist under a continuous tuning of experimental knobs, e.g., magnetic field and various gate voltages, as MBSs are predicted to be topologically protected from perturbations. On the contrary, $w_e$, $w_h$, and $\varepsilon_B$ of an ABS are anticipated to change when tuning experimental knobs. A natural question is whether $w_e=w_h=|w_{eh}|=0.5$ and $\varepsilon_B=0$ can simultaneously occur by accident. Our answer is negative, based on numerical simulations of the 1D Majorana nanowire model \cite{lutchyn2010majorana,oreg2010helical}:
$H_\textrm{wire}=\int_{0}^{L} dx\Psi^{\dag}(x){\cal H}\Psi(x)$ with ${\cal H}=\left[ p_{x}^{2}/2m^{\ast}-\mu+V_\textrm{pot}(x)-\alpha\sigma_{y} p_x/\hbar \right]  \tau_{z}+V_{Z}\sigma_{x}+\Delta\tau_{x}$.
Here $L$, $m^*$, $\mu$, and $p_x=-i\hbar\partial_x$, $\alpha$, $\Delta$ are the wire length, effective electron mass, chemical potential, momentum operator, Rashba spin-orbit coupling strength, and superconducting pairing potential, respectively. $V_Z=g_{\textrm{eff}}\mu_B B/2$ is the Zeeman energy induced by a magnetic field $B$, with $g_{\textrm{eff}}$ the effective Land\'{e} factor and $\mu_B$ the Bohr magneton. $V_\textrm{pot}(x)$ denotes the possible nonuniform electrostatic potential along the nanowire. With given parameters, diagonalizing $H_\textrm{wire}$ on a 1D lattice directly gives the energy spectrum and wave functions.

\begin{figure}[t!]
\centering
\includegraphics[width=\columnwidth]{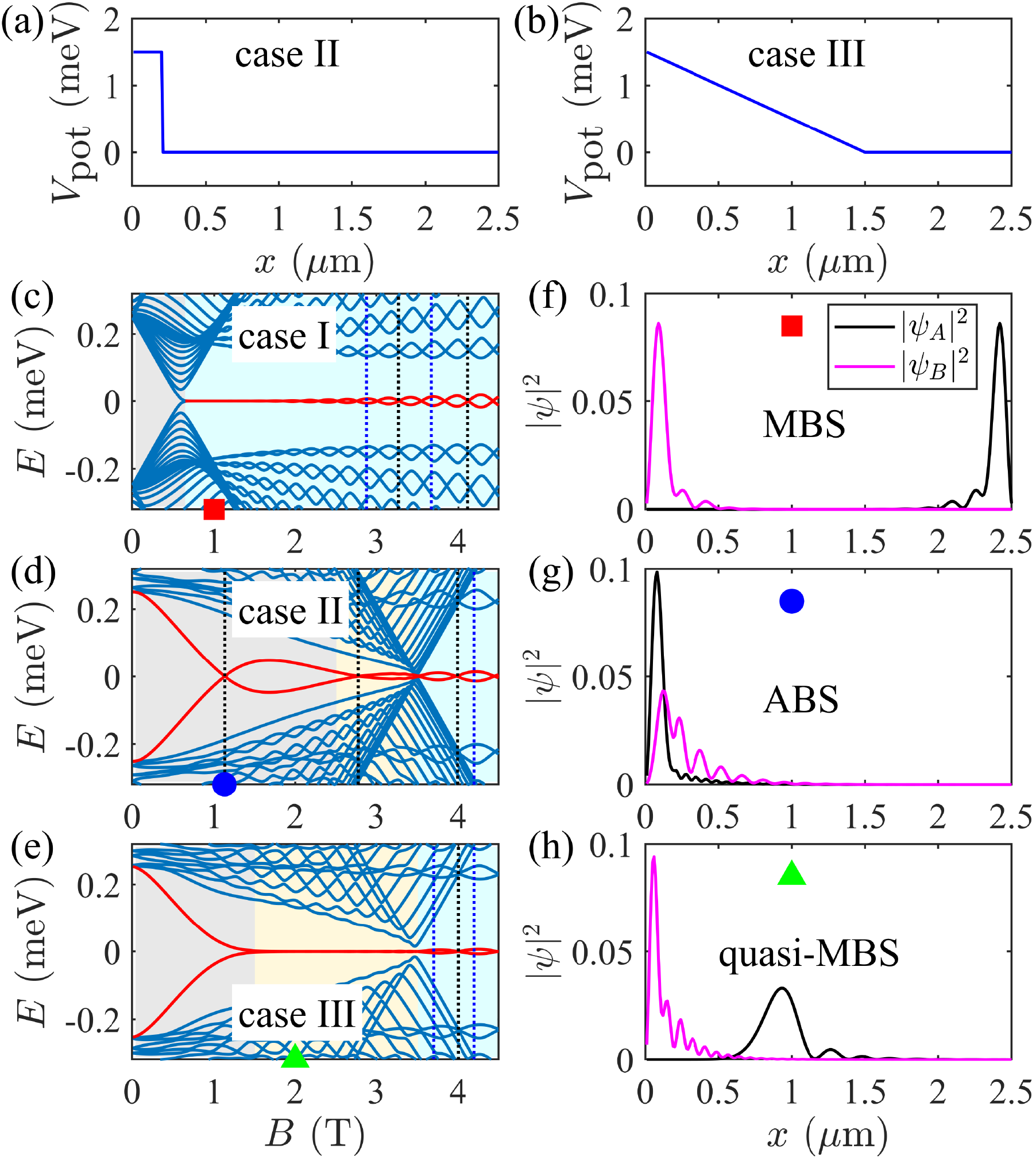}
\caption{[(a) and (b)] Profiles of two possible nonuniform electrostatic potential $V_\text{pot}(x)$ of a Majorana nanowire with a length of $L=2.5~\mu$m. [(c)--(e)] Energy spectra of the Majorana nanowire in three cases: (I) $\alpha=20$ meVnm and $\mu=V_\text{pot}(x)=0$; (II) $\alpha=30$ meVnm, $\mu=1.5$ meV, and $V_\text{pot}(x)$ shown in (a); and (III) $\alpha=40$ meVnm, $\mu=1.5$ meV, and $V_\text{pot}(x)$ shown in (b). The $B$-dependent LBSs traced in red are ABSs, quasi-MBSs, or MBSs, as highlighted by the background color of gray, yellow, or cyan, respectively. A few zero-energy and finite-energy LBSs are marked by black and blue dashed lines, respectively. [(f)--(h)] Spatial profiles of the squared modulus of the two wave functions in Majorana basis $\psi_A$ and $\psi_B$ [see Eqs.~\eqref{psia} and \eqref{psib}] of the LBSs at the magnetic fields indicated by the colored markers in (c)--(e). The spatial profiles of $|\psi_A|^2$ and $|\psi_B|^2$ shown in (f)--(h) are representative of MBSs, ABSs, and quasi-MBSs, respectively. Other parameters used for calculations are $m^\ast=0.026~m_e$, $\Delta=0.25$ meV, and $g_\textrm{eff}=15$. }\label{Fig:1Dmodel}
\end{figure}

\begin{figure}[t!]
\centering
\includegraphics[width=\columnwidth]{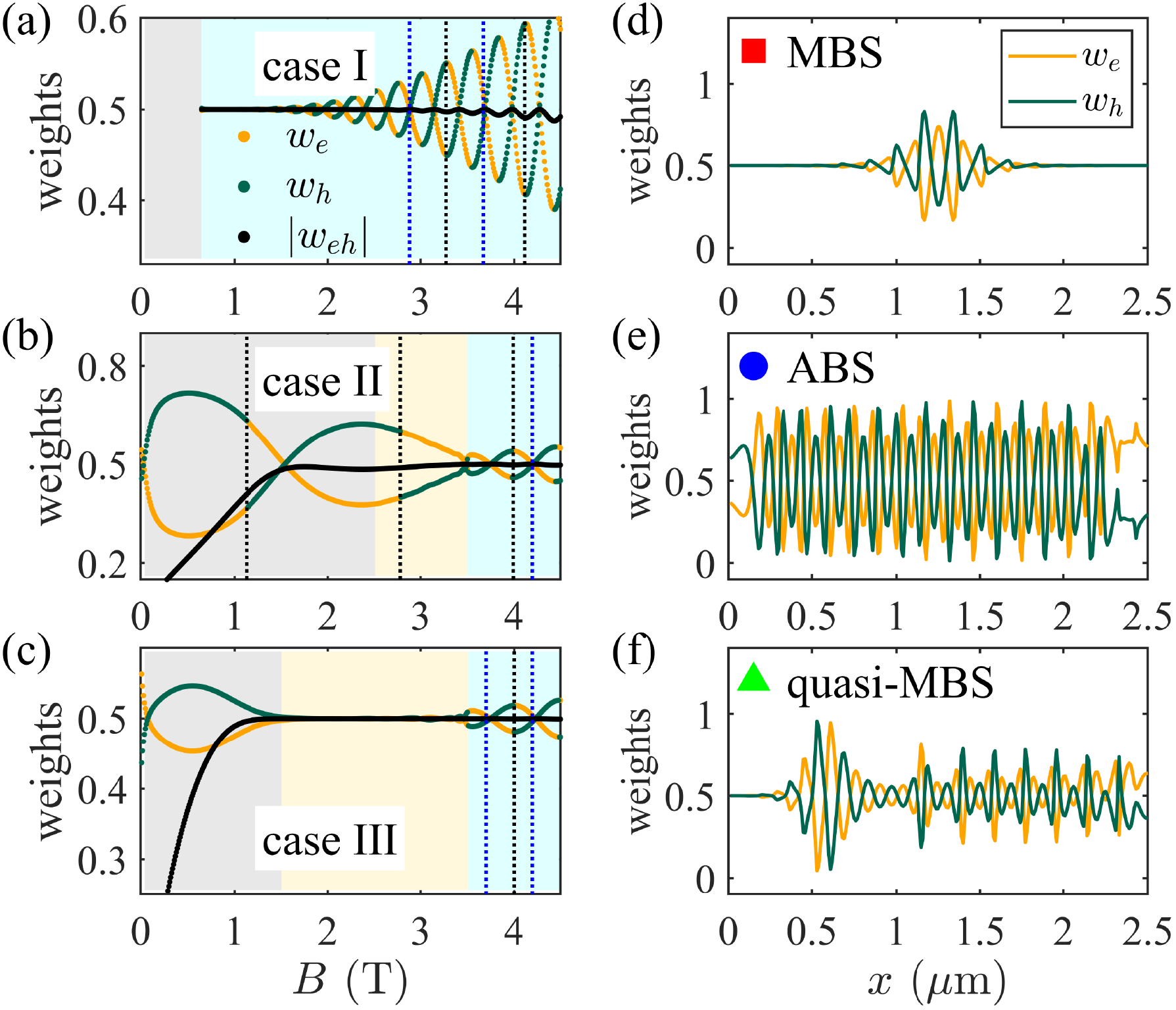}
\caption{[(a)--(c)] Normalized weights $w_e$, $w_h$, and $w_{eh}$ of the negative-energy branch LBS shown by the red lines in Fig.~\ref{Fig:1Dmodel}(c)--\ref{Fig:1Dmodel}(e). The weights are calculated at the leftmost site ($x=0$) of the nanowire. The same background colors and vertical dashed lines are also shown in Figs.~\ref{Fig:1Dmodel}(c)--\ref{Fig:1Dmodel}(e). In the cyan region of (a) and yellow region of (c), $w_e=w_h=|w_{eh}|=0.5$ persist over a sizable range of $B$, the corresponding LBSs are robust zero-energy MBSs and quasi-MBSs, respectively, shown in Figs.~\ref{Fig:1Dmodel}(c) and \ref{Fig:1Dmodel}(e). In Figs.~\ref{Fig:1Dmodel}(c)--\ref{Fig:1Dmodel}(e) and Figs.~\ref{Fig:distribution}(a)--\ref{Fig:distribution}(c), at some fine-tuned $B$ values the LBSs have $w_e=w_h=|w_{eh}|=0.5$ and $\varepsilon_B\ne 0$ (marked blue dashed lines) or $w_e\ne w_h\ne |w_{eh}|$ and $\varepsilon_B= 0$ (marked by black dashed lines). [(d)--(f)] Spatial distributions of the normalized weights $w_e$ and $w_h$ for the zero-energy MBS, ABS, and quasi-MBS indicated by the colored markers in Figs.~\ref{Fig:1Dmodel}(c)--\ref{Fig:1Dmodel}(e).}\label{Fig:distribution}
\end{figure}

We consider three nanowires with different electrostatic potentials: $V_\text{pot}(x)=0$ (case I), a steplike $V_\text{pot}(x)$ shown in Fig.~\ref{Fig:1Dmodel}(a) (case II), and a linear $V_\text{pot}(x)$ shown in Fig.~\ref{Fig:1Dmodel}(b) (case III). The numerically calculated $B$-dependent low-energy BdG spectra of these cases are presented in Figs.~\ref{Fig:1Dmodel}(c)-\ref{Fig:1Dmodel}(e), respectively. As $B$ increases, a topological phase transition occurs at the boundary of the cyan region (where $V_{Z}^c=\sqrt{\Delta^2+\textrm{max}\{[\mu-V_\text{pot}(x)]^2\}}$~\cite{zeng2022partially}), together with a gap close-and-reopen feature at $V_Z^c$ in the spectra. The red lines in the spectra label the lowest-energy BdG states (LBSs) that are generally well separated from the bulk states. For a given $B$, the negative $E$ in the red lines corresponds to the parameter $\varepsilon_B$ in Eq.~\eqref{H_ABS}. Following the red lines, LBSs with robust or accidental zero-energy can be found. The nature of a subgap bound state can be identified by visualizing the corresponding spatial profiles of the wave functions in the Majorana basis. Specifically, with the wave functions of the positive- and negative-energy LBS $\psi_{\pm}(x)=\{u_{\pm,\uparrow}(x),u_{\pm,\downarrow}(x),v_{\pm,\downarrow}(x),-v_{\pm,\uparrow}(x)\}$, we can construct the wave functions in the Majorana basis, i.e., the linear combinations~\cite{moore2018two}
\begin{eqnarray}
\psi_{A}(x)=\frac{1}{\sqrt{2}}[\psi_{+}(x)+\psi_{-}(x)],\label{psia}\\ \psi_{B}(x)=\frac{i}{\sqrt{2}}[\psi_{+}(x)-\psi_{-}(x)],\label{psib}
\end{eqnarray}
for the real Hamiltonian $H_\textrm{wire}$.
In Figs.~\ref{Fig:1Dmodel}(f)--\ref{Fig:1Dmodel}(h), We present the spatial profiles of $|\psi_A|^2$ and $|\psi_B|^2$ for the corresponding LBSs at the magnetic fields indicated by the colored markers in Figs.~\ref{Fig:1Dmodel}(c)--\ref{Fig:1Dmodel}(e), respectively. These profiles show the results of three representative cases: (i) $|\psi_A|^2$ and $|\psi_B|^2$ are well separated and localized at the opposite wire ends [Fig.~\ref{Fig:1Dmodel}(f)]; (ii) $|\psi_A|^2$ and $|\psi_B|^2$ extensively overlap and localize at the same wire end [Fig.~\ref{Fig:1Dmodel}(g)]; and (iii) $|\psi_A|^2$ and $|\psi_B|^2$ are spatially separated by a length that is larger than the peak widths, and only one of them is localized at the wire end [Fig.~\ref{Fig:1Dmodel}(h)]. The subgap bound states with these three characteristic spatial profiles of $|\psi_A|^2$ and $|\psi_B|^2$ are respectively referred to as MBS, ABS, and quasi-MBS \cite{vuik2019reproducing} (or partially separated ABS \cite{moore2018two,moore2018quantized,stanescu2019robust,cao2019decays}). In the spectra shown in Figs.~\ref{Fig:1Dmodel}(c)--\ref{Fig:1Dmodel}(e), we highlight the regions where the LBSs belong to ABSs, quasi-MBSs, or MBSs with a background color of gray, yellow, or cyan, respectively. Notice that the three regions are not strictly separated as the LBSs experience crossovers near the boundaries \cite{marra2022majorana}.

Now we proceed to discuss the electron-hole weights of ABSs, quasi-MBSs, and MBSs in Majorana nanowires. With the wave function $\psi_{-}(x)$, one can calculate the normalized weights
\begin{eqnarray}
w_{e}(x)&=&\frac{1}{\Lambda(x)}\sum_\sigma |u_{-,\sigma}(x)|^2,\\ w_h(x)&=&\frac{1}{\Lambda(x)}\sum_\sigma |v_{-,\sigma}(x)|^2,\\
w_{eh}(x)&=&\frac{1}{\Lambda(x)}\sum_\sigma u^\ast_{-,\sigma}(x)v_{-,\sigma}(x),
\end{eqnarray}
where we choose $\Lambda(x)=\sum_\sigma(|u_{-,\sigma}(x)|^2+|v_{-,\sigma}(x)|^2)$ as the normalization factor to reflect the relative electron and hole weights at all positions $x$. This renormalization is especially useful when the weights are small at some positions. In Figs.~\ref{Fig:distribution}(a)--\ref{Fig:distribution}(c), we show the $B$-dependent electron and hole weights at the leftmost site ($x=0$) of the nanowire, for the LBSs shown in red lines in Figs.~\ref{Fig:1Dmodel}(c)--\ref{Fig:1Dmodel}(e), respectively. These weights are relevant to the tunneling conductance measured in hybrid semiconductor-superconductor nanowire experiments (see, e.g., Refs.~\cite{deng2016Majorana,nichele2017scaling, gul2018ballistic,vaitiekenas2018effective,de2018electric,bommer2019spin,song2021large,wang2022observation}), in which the electrode is tunnel coupled to one end (say the left end) of the nanowire. In the cyan regions of Figs.~\ref{Fig:1Dmodel}(c) and \ref{Fig:distribution}(a) where MBSs exist and in the yellow regions of Figs.~\ref{Fig:1Dmodel}(e) and \ref{Fig:distribution}(c) where quasi-MBSs exist, one finds that $w_e=w_h=|w_{eh}|=0.5$ and $\varepsilon_B=0$ are robust over a sizable range of $B$.
In all three nanowire cases, the energy level of the MBS remarkably oscillates at large $B$ [the cyan regions of Figs.~\ref{Fig:1Dmodel}(c)--\ref{Fig:1Dmodel}(e)]. The oscillation arises from the inter-MBS hybridization of a finite-size nanowire~\cite{sarma2012splitting,rainis2013towards}. As shown in the cyan regions of Figs.~\ref{Fig:distribution}(a)--\ref{Fig:distribution}(c), this finite-size effect also leads to oscillations of the electron-hole weights of MBSs at large $B$.
Particularly, for certain values of $B$, MBSs have either (i) $w_e=w_h=|w_{eh}|=0.5$ and $\varepsilon_B\ne 0$ (marked by blue dotted lines) or (ii) $w_e\ne w_h\ne |w_{eh}|$ and $\varepsilon_B= 0$ (marked by black dotted lines).
MBSs of both cases, however, dramatically deviate from the aforementioned robust zero-energy MBSs with equal electron-hole weights. In nanowire case II, ABSs and quasi-MBSs with accidental zero-energy are shown to own $w_e\ne w_h\ne |w_{eh}|$, as indicated by the black dashed lines in the gray and yellow regions in Figs.~\ref{Fig:1Dmodel}(d) and \ref{Fig:distribution}(b). These results imply that as long as the zero-energy LBS persists owning both $\varepsilon_B=0$ and $w_e=w_h=0.5$ when continuously varying $B$, the value of $|w_{eh}|$ can be inferred to be 0.5 and the LBS can be identified as a robust zero-energy MBS or quasi-MBS, both of which are applicable in topological quantum computation \cite{vuik2019reproducing,zeng2020feasibility}.

As shown above, robust zero-energy MBSs and quasi-MBSs share an identical feature, i.e., $w_e=w_h=|w_{eh}|=0.5$, at the leftmost site of the nanowires, which have zero or nonuniform electrostatic potentials near the left wire end. However, their spatial distributions of the electron-hole weights are quite different, as shown in Figs.~\ref{Fig:distribution}(d)--\ref{Fig:distribution}(f), which correspond to the zero-energy LBSs indicated by the colored markers in Fig.~\ref{Fig:1Dmodel}. Specifically, in Fig.~\ref{Fig:distribution}(d), the zero-energy MBS owns $w_e=w_h=0.5$ everywhere except for in the middle of the nanowire, which is due to the finite-size effect for a nanowire with a moderate length. For an ideal MBS in the vortex core at the surface of an iron-based superconductor, the electron-hole weights have also been predicted to be equal anywhere away from the vortex core \cite{jiang2019quantum}. In Fig.~\ref{Fig:distribution}(f), for a zero-energy quasi-MBS, $w_e=w_h=0.5$ only shows up within the range $0<x<0.35~\mu$m, where $|\psi_A|^2$ and $|\psi_B|^2$ have a negligible overlap as shown in Fig.~\ref{Fig:1Dmodel}(h). By contrast, in Fig.~\ref{Fig:distribution}(e), for a zero-energy ABS, the associated $w_e$ and $w_h$ obviously unequal near the left wire end where most of its wave function weights are distributed [see Fig.~\ref{Fig:1Dmodel}(g)], and oscillate quickly elsewhere. We infer that the characteristic spatial distributions of electron-hole weights shown in Figs.~\ref{Fig:distribution}(d)--\ref{Fig:distribution}(f) can also exist in other candidate platforms supporting MBSs, such as topological insulator-superconductor heterostructures \cite{xu2015experimental,sun2016majorana} and iron-based superconductors \cite{zhang2018observation,wang2018evidence}. If the time-dependent current in such systems can be measured by ultrafast STM techniques \cite{loth2010measurement,cocker2013ultrafast}, the spatial distributions of electron-hole weights of the subgap bound states therein can be probed by our protocol and thus enables the identification of zero-energy MBSs.

\section{Summary}\label{summary}
To access the electron-hole weights $w_e$ and $w_h$ of an ABS in general superconducting systems, we have proposed to measure and fit the time-dependent local tunneling current induced by a steplike pulse bias. A direct application of our protocol is to detect MBSs in all candidate platforms where local tunneling spectroscopy can be measured. To elaborate this point for the Majorana nanowire system, we have numerically studied the energies, wave functions, and electron-hole weights of ABSs, quasi-MBSs, and MBSs therein. Interestingly, at the leftmost (or rightmost) site of the nanowire, where most of the wave functions of a bound state are distributed, the electron-hole weights are found unequal for accidental zero-energy ABSs and quasi-MBSs, but equal for zero-energy MBSs and quasi-MBSs that are
robust against the tuning of magnetic field. These zero-energy bound states also exhibit quite different spatial distributions of electron-hole weight over the nanowire. While accidental zero-energy ABSs own generally unequal electron-hole distributions along the nanowire, robust zero-energy MBSs and quasi-MBSs have equal electron-hole weights in the whole wire and a sizable region near one end of the wire, respectively. These remarkably distinct features of the electron-hole weights of different low-energy bound states in superconducting systems might be probed by our protocol with the recent advances in ultrafast electronic transports \cite{mciver2020light,torre2021colloquium,gutzler2021light}, thus is of timely importance in the experimental searching of topological MBSs.

\section{Acknowledgements}
This work was supported by the National Natural Science Foundation of China (Grants No.~12004040, No.~11974198, No.~92065206, and No.~12104043), the National Key Research and Development Program of China (Grant No. 2017YFA0303303), and Tsinghua University Initiative Scientific Research Program.

\begin{widetext}
\appendix
\section{Derivation of the time-dependent current mediated by an ABS}\label{appa}
Beyond the effective Hamiltonian \eqref{Htotal}--\eqref{Htunneling} studied in the main text, we begin with the general case where a metallic probe is coupled to a mesoscopic superconductor such as an atomic chain or ring, which can be described by a lattice model with $N$ sites
\begin{equation}
H_S=\sum_{i,j=1}^N\sum_{\sigma\sigma^{\prime}}(w_{ij\sigma\sigma^{\prime}}b^\dag_{i\sigma}b_{j\sigma^\prime} +\Delta_{ij\sigma\sigma^{\prime}}b_{i\sigma}b_{j\sigma^{\prime}}+\textrm{H.c.}).\label{H_S}
\end{equation}
General physics could be included by properly setting the parameters $w_{ij\sigma\sigma^\prime}$ and $\Delta_{ij\sigma\sigma^\prime}$. Specifically, the first term can describe the onsite energy, inter-site hopping, Zeeman splitting, spin-orbit coupling, and orbital effects of magnetic field, while the second term can describe superconducting pairing with an arbitrary symmetry. Correspondingly, the tunnel coupling between the probe and the superconductor is
\begin{equation}
H_T=\sum_{i=1}^N\sum_{k\sigma}(  \lambda_{i}c_{k\sigma}^{\dag}b_{i\sigma}+\textrm{H.c.}),
\end{equation}
where $\lambda_i$ represents the electron tunneling amplitude between the probe and site $i$. As we restrict to the subgap transport, i.e., the bias voltage smaller than the superconducting energy gap, the substrate supporting the mesoscopic superconductor is not modeled as its details are irrelevant.

The charge current flowing out of the probe, described by $H_P$ in the main text, can be calculated from the time evolution of the occupation number operator of the probe: $I(t)=-e\langle \dot{N}_{L}\rangle=-\frac{e}{i\hbar}\langle [N_P,H]\rangle$ with $N_P=\sum_{k\sigma}c_{k\sigma}^{\dag}c_{k\sigma}$. In the Nambu representation \cite{cuevas1996hamiltonian,sun1999photon} where the Hamiltonian, Green's functions (GFs), and self-energies are expressed in matrix form and denoted by boldface letters below, the charge current can be formally expressed as
\begin{eqnarray}
I^<(t)&=&\frac{e}{\hbar}\sum_{k}\textrm{Tr}[\mathbf{s}_{e}\mathbf{W}\mathbf{G}_{Sk}^{<}(  t,t)  -\mathbf{s}_{e}\mathbf{G}_{kS}^{<}(  t,t)  \mathbf{W}^{\dag}],\label{Ilesser1}\\
I^>(t)&=&\frac{e}{\hbar}\sum_{k}\textrm{Tr}[\mathbf{s}_{h}\mathbf{G}_{kS}^{>}(  t,t)  \mathbf{W}^{\dag}-\mathbf{s}_{h}\mathbf{W}\mathbf{G}_{Sk}^{>}(  t,t)],\label{Igreater1}
\end{eqnarray}
where $\mathbf{W}=[\mathbf{w}_1,...,\mathbf{w}_i,...,\mathbf{w}_N]$ is a $4\times 4N$ matrix with $\mathbf{w}_i=\textrm{diag}\{\lambda_i,\lambda_i,-\lambda_i^\ast,-\lambda_i^\ast\}$, $\mathbf{s}_{e}=\textrm{diag}\{1,1,0,0\}$, and $\mathbf{s}_{h}=\textrm{diag}\{0,0,1,1\}$. Note that $I^<(t)=I^>(t)$ and the superscript $<$ $(>)$ indicates that the current is expressed in terms of the lesser (greater) GF $\mathbf{G}^{<}_{Sk}$ ($\mathbf{G}^{>}_{kS}$). We derive both $I^<(t)$ and $I^>(t)$ for the reason that some interesting properties manifest themselves in the symmetric current formula $I(t)=\frac{1}{2}[I^<(t)+I^>(t)]$.

Using the equation-of-motion technique and analytic continuation rules \cite{haug2008quantum} one has
\begin{eqnarray}
\mathbf{G}_{Sk}^{<,>  }(  t,t)&=&\int dt_{1}\big[\mathbf{G}_{SS}^{r}(  t,t_{1})  \mathbf{W}^{\dag}\mathbf{g}_{kk}^{<,>}(  t_{1},t)+\mathbf{G}_{SS}^{<,>}(  t,t_{1})  \mathbf{W}^{\dag}\mathbf{g}_{kk}^{a}(t_{1},t)\big],\label{GSk}\\
\mathbf{G}_{kS}^{<,> }(  t,t)&=&\int dt_{1}\big[\mathbf{g}_{kk}^{r}(  t,t_{1})  \mathbf{WG}_{SS}^{<,> }(  t_{1},t)+\mathbf{g}_{kk}^{<,>}(  t,t_{1})  \mathbf{WG}_{SS}^{a}(  t_{1},t)\big],\label{GkS}
\end{eqnarray}
with the bare GFs of the probe
\begin{eqnarray}
\mathbf{g}_{kk}^{r,a}(t,t_1)&=&\mp i\Theta(  \pm t\mp t_1)[e^{i\phi(t,t_1)}\mathbf{s}_{e}+e^{-i\phi(t,t_1)}\mathbf{s}_{h}],\label{gkra}\\
\mathbf{g}_{kk}^{<,>}(t,t_1)&=&\pm i[f(\pm\varepsilon_{k})e^{i\phi(t,t_1)}\mathbf{s}_{e}+f(\mp\varepsilon_{k})e^{-i\phi(t,t_1)}\mathbf{s}_{h}],\label{gklg}
\end{eqnarray}
where $\Theta(t)$ is the Heaviside step function, $f(\varepsilon_{k})$ is the Fermi-Dirac distribution function, and $\phi_k(t,t_1)=\int_{t}^{t_1}dt'[\varepsilon_{k}+U(t')]$. Inserting Eqs.~\eqref{GSk}--\eqref{gklg} into Eqs.~\eqref{Ilesser1} and \eqref{Igreater1} yields
\begin{eqnarray}
I^<(  t)  &=&\frac{e}{\hbar}\int dt_{1}\textrm{Tr}\big[\mathbf{G}_{SS}^{r}(  t,t_{1})  \mathbf{\Sigma}_{e}^{<}(t_{1},t)  +\mathbf{G}_{SS}^{<}(  t,t_{1})  \mathbf{\Sigma}_{e}^{a}(  t_{1},t)-\mathbf{\Sigma}_{e}^{r}(t,t_{1})  \mathbf{G}_{SS}^{<}(  t_{1},t)  -\mathbf{\Sigma}_{e}^{<}(  t,t_{1})  \mathbf{G}_{SS}^{a}(  t_{1},t)\big], \label{Ilesser2}\\
I^>(t)  &=&\frac{e}{\hbar}\int dt_{1}\textrm{Tr}\big[\mathbf{\Sigma}_{h}^{r}(  t,t_{1})  \mathbf{G}_{SS}^{>}(t_{1},t)  +\mathbf{\Sigma}_{h}^{>}(  t,t_{1})  \mathbf{G}_{SS}^{a}(  t_{1},t)-\mathbf{G}_{SS}^{r}(  t,t_{1})\Sigma_{h}^{>}(  t_{1},t)  -\mathbf{G}_{SS}^{>}(t,t_{1})  \Sigma_{h}^{a}(  t_{1},t)\big], \label{Igreater2}
\end{eqnarray}
with
\begin{equation}
\mathbf{\Sigma}^{r,a,<,>}_{e(h)}(t,t_1)=\sum_{k}\mathbf{W}^{\dag}\mathbf{g}^{r,a,<,>}_{kk}(  t,t_1)\mathbf{s}_{e(h)}\mathbf{W},\label{ebse}
\end{equation}
where $\mathbf{G}_{SS}$ are the dressed GFs of the superconductor and $\mathbf{\Sigma}_{e(h)}$ are the embedded self-energies arising from the coupling between the probe and the superconductor. For simplicity, we shall drop the subscript of $\mathbf{G}_{SS}$ hereafter. The dressed retarded (advanced) GF $\mathbf{G}^{r,a}(t,t_1)$ and the lesser (greater) GF $\mathbf{G}^{<,>}(t,t_1)$ of the superconductor can be obtained through the Dyson and Keldysh equations, respectively, \cite{haug2008quantum}
\begin{eqnarray}
\mathbf{G}^{r,a}(t,t_1)&=&\mathbf{g}^{r,a}(  t,t_1)+\sum_{\eta=e,h}\iint dt^\prime dt^{\prime\prime}\mathbf{g}^{r,a}(  t,t^\prime)\mathbf{\Sigma}_\eta^{r,a}(  t^\prime,t^{\prime \prime})\mathbf{G}^{r,a}(  t^{\prime \prime},t_1),\label{dyson}\\
\mathbf{G}^{<,>}(t,t_1)&=&\sum_{\eta=e,h}\iint dt^\prime dt^{\prime \prime}\mathbf{G}^{r}(  t,t^\prime)\mathbf{\Sigma}_\eta^{<,>}(  t^\prime,t^{\prime \prime})\mathbf{G}^{a}(t^{\prime \prime},t_1),\label{keldysh}
\end{eqnarray}
with the bare GFs of the superconductor $\mathbf{g}^{r,a}(  t,t_1)=\mp i\Theta(\pm t\mp t_1)e^{-i\mathbf{H}_S(t-t_1)}$. It follows from Eqs.~\eqref{gkra} and \eqref{ebse} that
\begin{eqnarray}
\mathbf{\Sigma}_{e(  h)  }^{r,a}(t,t_1)=\int\frac{d\varepsilon}{2\pi}e^{-i\varepsilon(  t-t_1)  }[\mathbf{\Lambda}_{e(  h)  }( \varepsilon)  \mp\frac{i}{2}\mathbf{\Gamma}_{e(  h)  }( \varepsilon) ]  e^{\pm i\psi(t,t_1)},\notag\\
\end{eqnarray}
where $\psi(t,t_1)=\int_{t}^{t_1}dt'U(  t')$, $\mathbf{\Lambda}_{e(  h)  }(  \omega)  =\mathcal{P} \int\frac{d\varepsilon^{\prime}}{2\pi}\frac{\mathbf{\Gamma}_{e(  h)}( \varepsilon^{\prime})  }{\varepsilon-\varepsilon^{\prime}}$, and $\mathbf{\Gamma}_{e(  h)  }( \varepsilon)  =2\pi\sum_{k}\mathbf{W}^{\dag}\mathbf{s}_{e(  h)  }\mathbf{W}\delta(\varepsilon-\varepsilon_{k})$.

Solving integral Eqs.~\eqref{dyson} and \eqref{keldysh} with double time arguments is a nontrivial assignment \cite{zhu2005time,maciejko2006time}, but can be readily accomplished in the so-called wide-band limit \cite{wingreen1989inelastic}. To be specific, assuming a constant density of states $\rho$ of the probe gives $\mathbf{\Gamma}_{e(h)}(\varepsilon)\approx\mathbf{\Gamma}_{e(h)}=2\pi\rho\mathbf{W}^{\dag}\mathbf{s}_{e(h)}\mathbf{W}$, therefore,
\begin{eqnarray}
\mathbf{\Sigma}_{e(  h)  }^{r,a}(t,t_1)&=&\mp i\frac{1}{2}\mathbf{\Gamma}_{e(  h)}\delta(  t-t_1),\label{sra}\\
\mathbf{\Sigma}_{e(  h)  }^{<}(  t,t_1)&=&i\int \frac{d\varepsilon}{2\pi}f(  \varepsilon)  e^{i\int_{t}^{t_1}dt'[  \varepsilon\pm U(t') ]  }\mathbf{\Gamma}_{e(  h)  },\label{slesser}\\
\mathbf{\Sigma}_{e(  h)  }^{>}(  t,t_1)&=&-i\int \frac{d\varepsilon}{2\pi}f(  -\varepsilon)  e^{i\int_{t}^{t_1}dt'[  \varepsilon\pm U(t') ]  }\mathbf{\Gamma}_{e(  h)  }.\label{sgreater}
\end{eqnarray}
Inserting Eq.~\eqref{sra} into Eq.~\eqref{dyson} yields
\begin{equation}
\mathbf{G}^{r,a}( t,t_1)=\mathbf{G}^{r,a}( t-t_1)=\mp i\Theta(\pm t\mp t_1)e^{-i\mathbf{H}_S^{r,a}(t-t_1)},\label{Grat}
\end{equation}
and its Fourier transformation is
\begin{equation}
\mathbf{G}^{r,a}(\varepsilon)=\int dt \mathbf{G}^{r,a}( t-t_1)e^{i\varepsilon(t-t_1)}=(\varepsilon-\mathbf{H}_S^{r,a})^{-1},\label{Grae}
\end{equation}
where $\mathbf{H}_S^{r,a}=\mathbf{H}_{S}\mp \frac{i}{2}(\mathbf{\Gamma}_{e}+\mathbf{\Gamma}_{h})$ and ${\mathbf{H}_S^{a}}^\dag=\mathbf{H}_S^{r}$. While the Hamiltonian matrix $\mathbf{H}_S$ is Hermitian, $\mathbf{H}_S^{r,a}$ is non-Hermitian owning $4N$ complex eigen-energies $E^{r,a}_m$ ($m=1,2,...,4N$), with $E^{a\ast}_m=E^{r}_m$ and $\textrm{Im}E^{r}_m\leq 0$. The $\delta$ function in $\mathbf{\Sigma}_{e(h)}^{r,a}(t,t_1)$ renders the involved lesser and greater GFs in Eqs.~\eqref{Ilesser2} and \eqref{Igreater2}, respectively, being with equal time arguments
\begin{equation}
\mathbf{G}^{<,>}(  t,t)=\pm i\sum_{\eta=e,h}\int \frac{d\varepsilon}{2\pi}f(\pm\varepsilon)  \mathbf{A}_{\eta}^{r}(  \varepsilon,t)\mathbf{\Gamma}_{\eta}\mathbf{A}_{\eta}^{a}(  \varepsilon,t),
\end{equation}
with
\begin{eqnarray}
\mathbf{A}_{\eta}^{r}(  \varepsilon,t)&=&\int dt_{1}\mathbf{G}^{r}(t- t_{1})e^{ i\int_{t_{1}}^{t}dt^{\prime}[  \varepsilon+s_\eta U(  t^{\prime}) ]  },\label{Ar}\\
\mathbf{A}_{\eta}^{a}(  \varepsilon,t)&=&\int dt_{1}\mathbf{G}^{a}(t_{1}-t)e^{- i\int_{t_{1}}^{t}dt^{\prime}[  \varepsilon+s_\eta U(  t^{\prime}) ]  },\label{Aa}
\end{eqnarray}
where $s_\eta=1$ for $\eta=e$ and $s_\eta=-1$ for $\eta=h$. Clearly, in the time-independent case $\mathbf{A}_{\eta}^{r,a}(\varepsilon,t)$ is the Fourier transform of $\mathbf{G}^{r,a}(\pm t\mp t_{1})$.

Inserting Eqs.~\eqref{sra}--\eqref{Aa} into Eqs.~\eqref{Ilesser2} and \eqref{Igreater2} yields the charge current induced by an arbitrary time-dependent bias voltage
\begin{eqnarray}
I^<(  t)&=&\frac{e}{h}\int d\varepsilon f(  \varepsilon) \textrm{Tr}\Big\{  i\mathbf{\Gamma}_{e}[\mathbf{A}_{e}^{r}(  \varepsilon,t)  -\mathbf{A}_{e}^{a}(\varepsilon,t)  ]  -\mathbf{\Gamma}_{e}\sum_{\eta=e,h}\mathbf{A}_{\eta}^{r}(  \varepsilon,t)\mathbf{\Gamma}_{\eta}\mathbf{A}_{\eta}^{a}(  \varepsilon,t)\Big\},\\
I^>(  t)&=&\frac{e}{h}\int d\varepsilon f(-\varepsilon) \textrm{Tr}\Big\{  i\mathbf{\Gamma}_{h}[\mathbf{A}_{h}^{r}(  \varepsilon,t)  -\mathbf{A}_{h}^{a}(\varepsilon,t)  ]-\mathbf{\Gamma}_{h}\sum_{\eta=e,h}\mathbf{A}_{\eta}^{r}(  \varepsilon,t)\mathbf{\Gamma}_{\eta}\mathbf{A}_{\eta}^{a}(  \varepsilon,t)\Big\}.
\end{eqnarray}
The symmetric current formula is
\begin{eqnarray}
I(t)&=&\frac{1}{2}[I^<(  t)+I^>(t)]\notag\\
&=&\frac{e}{2h}\int d\varepsilon  \textrm{Tr}\Big\{  i\sum_{\eta=e,h}f(  s_\eta\varepsilon)\mathbf{\Gamma}_{\eta}[\mathbf{A}_{\eta}^{r}(  \varepsilon,t)  -\mathbf{A}_{\eta}^{a}(\varepsilon,t)  ]  -[\mathbf{\Gamma}_{e}f( \varepsilon)+\mathbf{\Gamma}_{h}f(-\varepsilon)]\sum_{\eta=e,h}\mathbf{A}_{\eta}^{r}(  \varepsilon,t)\mathbf{\Gamma}_{\eta}\mathbf{A}_{\eta}^{a}(  \varepsilon,t)\Big\}.\label{It_symmetric}
\end{eqnarray}
This is a formally exact current formula applicable to systems comprising a metallic probe, applied with an arbitrary time-dependent bias voltage, coupled to general mesoscopic superconductors described by the lattice model Eq.~(\ref{H_S}).

Under the steplike pulse bias, $U(t)=\Theta(t)V$, Eq.~\eqref{Ar} is recast as
\begin{eqnarray}
\mathbf{A}_{\eta}^{r}(  \varepsilon,t)&&=\Theta(-t)\mathbf{G}^{r}(\varepsilon)+\Theta(t)\int dt_{1}\mathbf{G}^{r}(  t-t_{1})  e^{i(  \varepsilon+s_\eta V)  t}[\Theta(  t_{1})  e^{-i(  \varepsilon+s_\eta V)  t_{1}}+\Theta(  -t_{1})  e^{-i\varepsilon t_{1}}\big]  \notag\\
&& =\Theta(-t)\mathbf{G}^{r}(\varepsilon)+\Theta(t)\big\{\mathbf{G}^{r}(  \varepsilon+s_\eta V)+e^{i(\varepsilon+s_\eta V)  t}\int dt_{1}\Theta(  -t_{1})\mathbf{G}^{r}(  t-t_{1})[ -e^{-i(  \varepsilon+s_\eta V)  t_{1}}+e^{-i\varepsilon t_{1}}]\big\}.\label{B1}
\end{eqnarray}
With Eqs.~\eqref{Grat} and \eqref{Grae}, the integral in Eq.~\eqref{B1} is analytically evaluated as
\begin{eqnarray}
ie^{-i\mathbf{H}_{S}^{r}t}\int_{-\infty}^{0}dt_{1}[  e^{-i(  \varepsilon+s_\eta V-\mathbf{H}_{S}^{r})  t_{1}}-e^{-i(\varepsilon-\mathbf{H}_{S}^{r})  t_{1}}]&=&e^{-i\mathbf{H}_{S}^{r}t}\left(\frac{1}{\varepsilon-\mathbf{H}_{S}^{r}}-\frac{1}{\varepsilon+s_\eta V-\mathbf{H}_{S}^{r}}\right)\notag\\
&=&s_\eta Ve^{-i\mathbf{H}_{S}^{r}t}\mathbf{G}^{r}(  \varepsilon)  \mathbf{G}^{r}(  \varepsilon+s_\eta V),\label{B2}
\end{eqnarray}
where the fact $e^{-i\mathbf{H}^r_S \times (-\infty)}=0$ is used. Inserting Eq.~(\ref{B2}) into Eq.~(\ref{B1}) yields
\begin{equation}
\mathbf{A}_{\eta}^{r}(  \varepsilon,t)=\Theta(-t)\mathbf{G}^{r}(\varepsilon)+\Theta(t)\big[\mathbf{G}^{r}(\varepsilon+s_\eta V)+s_\eta Ve^{i(  \varepsilon+s_\eta V-\mathbf{H}_S^{r})t}\mathbf{G}^{r}(  \varepsilon)  \mathbf{G}^{r}(\varepsilon+s_\eta V)\big].\label{B3}
\end{equation}
The advanced function $\mathbf{A}_{\eta}^{a}(  \varepsilon,t)$ can be obtained by the relation $\mathbf{A}_{\eta}^{a}(  \varepsilon,t)=[\mathbf{A}_{\eta}^{r}(  \varepsilon,t)]^\dag$. As $\mathbf{G}^{r,a}(  \varepsilon)=(\varepsilon-\mathbf{H}_S^{r,a})  ^{-1}$, we have $s_\eta V \mathbf{G}^{r,a}(  \varepsilon)  \mathbf{G}^{r,a}(\varepsilon+s_\eta V)=\mathbf{G}^{r,a}(  \varepsilon)-  \mathbf{G}^{r,a}(\varepsilon+s_\eta V)$, thus it can be readily checked that the functions $\mathbf{A}_{\eta}^{r,a}(  \varepsilon,t)$ are continuous at $t=0$. Inserting Eq.~\eqref{B3} into Eq.~\eqref{It_symmetric} we arrive at
\begin{eqnarray}
\hspace{-0.5cm}I(t)&=&\Theta(t)[I_{\textrm{st}}+I_\textrm{trans}(  t) ],\label{B5}
\end{eqnarray}
with the steady-state current
\begin{eqnarray}
I_{\textrm{st}}&=&\frac{e}{2h}\sum_{\eta=e,h}\int d\varepsilon[  f(\varepsilon-V)  -f(  \varepsilon+V) ]\textrm{Tr}\big[\mathbf{\Gamma}_{\eta}\mathbf{G}^{r}(  \varepsilon)\mathbf{\Gamma}_{\bar\eta}\mathbf{G}^{a}(  \varepsilon)\big],\label{B7}
\end{eqnarray}
and the transient current $I_\textrm{trans}(  t)=I_1(  t)+I_2(  t)$ with
\begin{eqnarray}
I_{1}(t)&=&\frac{e}{2h}\sum_{\eta=e,h}\int d\varepsilon f(s_\eta\varepsilon)\textrm{Tr}\Big[is_\eta V\mathbf{\Gamma}_{\eta}e^{i(  \varepsilon+ s_\eta V-\mathbf{H}_{S}^{r})  t}\mathbf{G}^{r}(  \varepsilon)  \mathbf{G}^{r}(\varepsilon+s_\eta V) +\textrm{H.c.}\Big],\label{B8}\\
I_{2}(  t)&=&\frac{e}{2h}\sum_{\eta=e,h}\int d\varepsilon  \textrm{Tr}\Big\{\Big[-s_\eta V\big[f(\varepsilon)\mathbf{\Gamma}_{e}+f(-\varepsilon)\mathbf{\Gamma}_{h}\big]e^{i(  \varepsilon+s_\eta V-\mathbf{H}_{S}^{r})  t}\mathbf{G}^{r}(  \varepsilon)  \mathbf{G}^{r}(  \varepsilon+s_\eta V)  \mathbf{\Gamma}_{\eta}\mathbf{G}^{a}(  \varepsilon+s_\eta V)+\textrm{H.c.}\Big]\notag\\
&&\hspace{3cm}-V^{2}[f(\varepsilon)\mathbf{\Gamma}_{e}+f(-\varepsilon)\mathbf{\Gamma}_{h}]e^{-i\mathbf{H}_{S}^{r}t}\mathbf{G}^{r}(\varepsilon)  \mathbf{G}^{r}(  \varepsilon+s_\eta V)\mathbf{\Gamma}_{\eta}\mathbf{G}^{a}(  \varepsilon+s_\eta V)\mathbf{G}^{a}(  \varepsilon)    e^{i\mathbf{H}_{S}^{a}t}\Big\}.\label{B10}
\end{eqnarray}

In numerical calculations, the trace operations in Eqs.~(\ref{B7})--(\ref{B10}) can be implemented in the eigenbasis of the Hamiltonian $\mathbf{H}^{r,a}_S$. Specifically, there exists a similarity transformation ${\mathbf{U}^{r,a}}^{-1}\mathbf{H}_S^{r,a} \mathbf{U}^{r,a}=\mathbf{E}^{r,a}$, where $\mathbf{E}^{r,a}$ is a diagonal matrix comprising the eigen-energies $E^{r,a}_m$ ($m=1,2,...,4N$) of $\mathbf{H}^{r,a}_S$ and $\mathbf{U}^{r,a}$ consists of the eigen-vectors of $\mathbf{H}^{r,a}_S$. Inserting the identity $\mathbf{U}^{r,a}{\mathbf{U}^{r,a}}^{-1}=\mathbf{1}$ into proper positions in Eq.~\eqref{B7} yields
\begin{eqnarray}
I_{\textrm{st}}&&=\frac{e}{2h}\sum_{\eta=e,h}\int d\varepsilon [f(\varepsilon-V)-f(\varepsilon+V) ]\textrm{Tr}\Big[\mathbf{\Gamma}_{\eta}\mathbf{U}^{r}{\mathbf{U}^{r}}^{-1}\mathbf{G}^{r}(  \varepsilon)\mathbf{U}^{r}{\mathbf{U}^{r}}^{-1} \mathbf{\Gamma}_{\bar\eta}\mathbf{U}^{a}{\mathbf{U}^{a}}^{-1} \mathbf{G}^{a}(\varepsilon)\mathbf{U}^{a}{\mathbf{U}^{a}}^{-1}\Big]\notag\\
&&=\frac{e}{2h}\sum_{\eta=e,h}\int d\varepsilon [f(\varepsilon-V)-f(\varepsilon+V) ]\sum_{mn}\frac{\mathbf{\Gamma}^{ar}_{\eta;mn}\mathbf{\Gamma}^{ra}_{\bar\eta;nm}}{(  \varepsilon-E_{n}^r)(\varepsilon-E_m^a)},\label{C1}
\end{eqnarray}
Similarly, one can obtain
\begin{eqnarray}
I_{1}(t)&&=\frac{e}{2h}\sum_{\eta=e,h}\int d\varepsilon f(s_\eta\varepsilon) \sum_{n}\Bigg[\frac{i s_\eta V\mathbf{\Gamma}^{rr}_{\eta;nn}e^{i(\varepsilon+ s_\eta V-E_{n}^{r})  t}}{(\varepsilon-E_{n}^{r})(\varepsilon+s_\eta V-E_{n}^{r})} +\frac{-i s_\eta V\mathbf{\Gamma}^{aa}_{\eta;nn}e^{-i(\varepsilon+ s_\eta V-E_{n}^{a})  t}}{(\varepsilon-E_{n}^{a})(\varepsilon+s_\eta V-E_{n}^{a})}\Bigg],\label{C2}\\
I_{2}(  t)&&=\frac{e}{2h}\sum_{\eta=e,h}\int d\varepsilon \sum_{mn}\big[f(\varepsilon)\mathbf{\Gamma}_{e;mn}^{ar}+f(-\varepsilon)\mathbf{\Gamma}_{h;mn}^{ar}\big]\mathbf{\Gamma}^{ra}_{\eta;nm}\bigg[\frac{-s_\eta V e^{i(  \varepsilon+s_\eta V-E_{n}^{r})  t}}{(\varepsilon-E_{n}^{r})(\varepsilon+s_\eta V-E_{n}^{r})(\varepsilon+s_\eta V-E_{m}^{a})}\notag\\
&&+\frac{-s_\eta V e^{-i(  \varepsilon+s_\eta V-E_{m}^{a})  t}}{(\varepsilon-E_{m}^{a})(\varepsilon+s_\eta V-E_{m}^{a})(\varepsilon+s_\eta V-E_{n}^{r})}+\frac{-V^2 e^{-i(E_{n}^{r}-E_{m}^{a})t}}{(\varepsilon-E_{n}^{r})(\varepsilon+s_\eta V-E_{n}^{r})(\varepsilon+s_\eta V-E_{m}^{a})(\varepsilon-E_{m}^{a})}\bigg],\label{C4}
\end{eqnarray}
where $\mathbf{\Gamma}^{rr}_{\eta}={\mathbf{U}^{r}}^{-1}\mathbf{\Gamma}_{\eta}\mathbf{U}^{r}$, $\mathbf{\Gamma}^{ra}_{\eta}={\mathbf{U}^{r}}^{-1}\mathbf{\Gamma}_{\eta}\mathbf{U}^{a}$, and $\mathbf{\Gamma}^{ar}_{\eta}={\mathbf{U}^{a}}^{-1}\mathbf{\Gamma}_{\eta}\mathbf{U}^{r}$. By contour integration, it is readily to verify that the integration of the terms within the brackets in Eq.~\eqref{C4} over $\varepsilon$ are zeros, therefore, noting that $f(\varepsilon)+f(-\varepsilon)=1$, Eq.~\eqref{C4} can be recast as
\begin{eqnarray}
I_{2}(  t)&&=\frac{e}{2h}\sum_{\eta=e,h}\int d\varepsilon f(\varepsilon)\sum_{mn}(\mathbf{\Gamma}_{e;mn}^{ar}-\mathbf{\Gamma}_{h;mn}^{ar})\mathbf{\Gamma}^{ra}_{\eta;nm}\bigg[\frac{-s_\eta V e^{i(  \varepsilon+s_\eta V-E_{n}^{r})  t}}{(\varepsilon-E_{n}^{r})(\varepsilon+s_\eta V-E_{n}^{r})(\varepsilon+s_\eta V-E_{m}^{a})}\notag\\
&&+\frac{-s_\eta V e^{-i(  \varepsilon+s_\eta V-E_{m}^{a})  t}}{(\varepsilon-E_{m}^{a})(\varepsilon+s_\eta V-E_{m}^{a})(\varepsilon+s_\eta V-E_{n}^{r})}+\frac{-V^2 e^{-i(E_{n}^{r}-E_{m}^{a})t}}{(\varepsilon-E_{n}^{r})(\varepsilon+s_\eta V-E_{n}^{r})(\varepsilon+s_\eta V-E_{m}^{a})(\varepsilon-E_{m}^{a})}\bigg],\label{C6}
\end{eqnarray}
which can be reexpressed with a trace operator as
\begin{eqnarray}
I_{2}(  t)&=&\frac{e}{2h}\sum_{\eta=e,h}\int d\varepsilon f(\varepsilon) \textrm{Tr}\Big\{\Big[-s_\eta V (\mathbf{\Gamma}_{e}-\mathbf{\Gamma}_{h})e^{i(  \varepsilon+s_\eta V-\mathbf{H}_{S}^{r})  t}\mathbf{G}^{r}(  \varepsilon)  \mathbf{G}^{r}(  \varepsilon+s_\eta V)  \mathbf{\Gamma}_{\eta}\mathbf{G}^{a}(  \varepsilon+s_\eta V)+\textrm{H.c.}\Big]\notag\\
&&\hspace{3.8cm}-V^{2}(\mathbf{\Gamma}_{e}-\mathbf{\Gamma}_{h})e^{-i\mathbf{H}_{S}^{r}t}\mathbf{G}^{r}(\varepsilon)  \mathbf{G}^{r}(  \varepsilon+s_\eta V)\mathbf{\Gamma}_{\eta}\mathbf{G}^{a}(  \varepsilon+s_\eta V)\mathbf{G}^{a}(  \varepsilon)    e^{i\mathbf{H}_{S}^{a}t}\Big\}.\label{C8}
\end{eqnarray}

Equations \eqref{B7}, \eqref{B8}, and \eqref{C8} are applicable to the effective Hamiltonian \eqref{Htotal}--\eqref{Htunneling} and they are presented as Eqs.~\eqref{Ist}, \eqref{I1}, and \eqref{I2} in the main text, together with the corresponding expressions of $\mathbf{\Gamma}_e$, $\mathbf{\Gamma}_h$, $\mathbf{H}^{r(a)}_S$, and $\mathbf{G}^{r,a}(\varepsilon)$.

\section{Fourier spectrum of the transient current mediated by an ABS}\label{appc}
The Fourier spectrum of the transient current is $I(\omega)=I_1(\omega)+I_{2}(\omega)$ with
\begin{equation}
I_{1/2}(\omega)=\int_{0}^\infty dt I_{1/2}(t)e^{i\omega t}.\label{C10}
\end{equation}
Inserting Eqs.~(\ref{C2}) and (\ref{C6}) into Eq.~(\ref{C10}), respectively, and integrating analytically over the time argument yields
\begin{eqnarray}
I_{1}(\omega)&&=\frac{e}{2h}\sum_{\eta=e,h}\sum_{n}\int d\varepsilon f(s_\eta\varepsilon) \Big[ \frac{-s_\eta V\mathbf{\Gamma}_{\eta;nn}^{rr}}{(\varepsilon+\omega+s_\eta V-E_{n}^{r}) (  \varepsilon-E_{n}^{r})(  \varepsilon+s_\eta V-E_{n}^{r})  }\notag\\
&&+\frac{-s_\eta V\mathbf{\Gamma}_{\eta;nn}^{aa}}{(\varepsilon-\omega+s_\eta V-E_{n}^{a}) (  \varepsilon-E_{n}^{a})(  \varepsilon+s_\eta V-E_{n}^{a})  }\Big],\label{C11}\\
I_{2}(  \omega)&&=\frac{e}{2h}\sum_{\eta=e,h}\sum_{mn}\int d\varepsilon f(  \varepsilon)[\mathbf{\Gamma}_{e;mn}^{ar}-\mathbf{\Gamma}_{h;mn}^{ar}]\mathbf{\Gamma}_{\eta;nm}^{ra} \Big[ \frac{-i s_{\eta}V }{(  \varepsilon+\omega+s_{\eta}V-E_{n}^{r})(  \varepsilon-E_{n}^{r})(  \varepsilon+s_{\eta}V-E_{n}^{r}) (\varepsilon+s_{\eta}V-E_{m}^{a})   }\notag\\
&&+\frac{i s_{\eta}V }{(  \varepsilon-\omega+s_{\eta}V-E_{m}^{a})(  \varepsilon-E_{m}^{a})(  \varepsilon+s_{\eta}V-E_{m}^{a}) (\varepsilon+s_{\eta}V-E_{n}^{r})   }\notag\\
&&+\frac{-iV^{2}}{(\omega-E_n^r+E_m^a)(  \varepsilon-E_{n}^{r}) (  \varepsilon+s_{\eta}V-E_{n}^{r})  (  \varepsilon-E_{m}^{a}) (\varepsilon+s_{\eta}V-E_{m}^{a})  }\Big].\label{C13}
\end{eqnarray}
At zero temperature, the integrals over the energy argument in Eqs.~\eqref{C11} and \eqref{C13} can be analytically evaluated,
\begin{eqnarray}
I_{1}(\omega)&&  =\frac{e}{2h}\sum_{\eta=e,h}\sum_{n}\Bigg\{s_\eta \mathbf{\Gamma}_{\eta;nn}^{rr}\bigg[-\frac{\ln E_{n}^{r}-\ln(E_{n}^{r}-\omega-s_\eta V)  }{\omega+s_\eta V}+\frac{\ln(  E_{n}^{r}-s_\eta V)-\ln(  E_{n}^{r}-s_\eta V-\omega)}{\omega}\bigg]\notag\\
&&+s_\eta \mathbf{\Gamma}_{\eta;nn}^{aa}\bigg[  -\frac{\ln E_{n}^{a}-\ln(E_{n}^{a}+\omega-s_\eta V)  }{-\omega+s_\eta V}+\frac{\ln(  E_{n}^{a}-s_\eta V)-\ln(  E_{n}^{a}-s_\eta V+\omega)}{-\omega}\bigg]\Bigg\},\label{I1omega}\\
I_{2}(  \omega)=&&\frac{ie}{2h}\sum_{\eta=e,h}\sum_{mn}\frac{\mathbf{\Gamma}_{e;mn}^{ar}-\mathbf{\Gamma}_{h;mn}^{ar}}{\omega-(E_{n}^{r}-E_{m}^{a})}\mathbf{\Gamma}_{\eta;nm}^{ra}\Big[\frac{\ln E_{n}^{r}-\ln(  E_{n}^{r}-\omega-s_\eta V)  }{\omega+s_\eta V}-\frac{\ln(  E_{n}^{r}-s_\eta V)  -\ln(  E_{n}^{r}-s_\eta V-\omega)  }{\omega}\notag\\
&&+\frac{\ln E_{m}^{a}-\ln(  E_{m}^{a}+\omega-s_\eta V)  }{-\omega+s_\eta V}-\frac{\ln(  E_{m}^{a}-s_\eta V)  -\ln(  E_{m}^{a}-s_\eta V+\omega)  }{-\omega}\notag\\
&&+\frac{  \ln(  E_{n}^{r}-s_\eta V) -\ln E_{n}^{r}-  \ln(  E_{m}^{a}-s_\eta V) +\ln E_{m}^{a}  }{E_{n}^{r}-E_{m}^{a}}\Big].\label{I2omega}
\end{eqnarray}
It is clear that characteristic peaks emerge at $\omega=|V\pm \textrm{Re}E_n^r|$ in $I_1(\omega)$ and $I_2(\omega)$ with logarithmic lineshapes, and at $\omega=|\textrm{Re}E_n^r-\textrm{Re}E_m^a|$ in $I_2(\omega)$ with Lorentzian lineshapes. Equations \eqref{I1omega} and \eqref{I2omega} reveal three transport processes contributing to the transient current. (i) The coherent electron transitions between the probe and the $n$th eigenstate of the dressed superconductor, i.e., $\textrm{H}^r_S$, result in an oscillating decaying current component with frequencies $|\omega|=|V \pm \textrm{Re}E_n^r|$ and a damping rate $\textrm{Im}E_n^r$. (ii) The coherent electron transitions between the $m$th and $n$th eigenstates of the dressed superconductor result in an oscillating decaying current component with frequencies $|\omega|=|\textrm{Re}E_n^r-\textrm{Re}E_m^a|$ and a damping rate $|\textrm{Im}E_n^r-\textrm{Im}E_m^a|$. (iii) After the sudden switch-on of the bias, the time evolution of the occupation probability of the $n$th eigenstate of the dressed superconductor results in an exponentially decaying current component with a damping rate $2|\textrm{Im}E_n^r|$, which is just captured by our RE analysis in the main text.

For the effective Hamiltonian $H$ \eqref{Htotal}--\eqref{Htunneling} studied in the main text, $m,n=\{1,2\}$ and process (ii) is absent due to that $\mathbf{\Gamma}^{ar}_{e;12}=\mathbf{\Gamma}^{ar}_{h;12}$ and $\mathbf{\Gamma}^{ar}_{e;21}=\mathbf{\Gamma}^{ar}_{h;21}$ render the associated coefficients in Eq.~\eqref{I2omega} being zeros.

\end{widetext}

%\bibliographystyle{apsrev4-1-title}
%\bibliographystyle{apsrev4-1}
%\bibliography{refs-Majorana}


\end{document}